\documentclass[11pt,a4paper]{article}
\pdfoutput=1

\usepackage[utf8]{inputenc}
\usepackage{a4wide}
\usepackage{amsmath}
\usepackage{cite}
\usepackage{xcolor}
\usepackage{amssymb}
\usepackage{amsfonts}
\usepackage{booktabs}
\usepackage{makecell}
\usepackage{multirow}
\usepackage{pdflscape}
\usepackage{textcomp}
\usepackage{gensymb}
\usepackage{bigstrut}
\usepackage{array}
\usepackage{enumerate}
\usepackage[small,bf]{caption}
\usepackage{graphicx}
\usepackage{hyperref}
\usepackage{mathbbol}
\usepackage{appendix}
\usepackage{verbatim}
\usepackage{geometry}
\usepackage{makecell}
\usepackage{listings}
\usepackage{mathtools}
\usepackage{dsfont}
\usepackage{accents}
\usepackage{longtable}
\usepackage{colortbl}

\usepackage{floatpag}

\usepackage[capitalise]{cleveref} 

\makeatletter
\newcommand {\crefext}[2]{\csname cref@#1@format\endcsname{#2}{}{}}
\newcommand {\Crefext}[2]{\csname Cref@#1@format\endcsname{#2}{}{}}
\makeatother

\DeclareMathOperator{\im}{Im}
\DeclareMathOperator{\re}{Re}

\newcommand*{\belowrulesepcolor}[1]{%
  \noalign{%
    \kern-\belowrulesep
    \begingroup
      \color{#1}%
      \hrule height\belowrulesep
    \endgroup
  }%
}
\newcommand*{\aboverulesepcolor}[1]{%
  \noalign{%
    \begingroup
      \color{#1}%
      \hrule height\aboverulesep
    \endgroup
    \kern-\aboverulesep
  }%
}

\allowdisplaybreaks
\numberwithin{equation}{section}

\begin{document}

\begin{titlepage}

\vspace*{-15mm}
\begin{flushright}
SISSA 22/2021/FISI\\
IPMU21-0088\\
CFTP/21-015\\
IPhT-T22/005
\end{flushright}
\vspace*{5mm}

\begin{center}
{\bf\LARGE {Modular Flavour Symmetries\\[2mm] and Modulus Stabilisation}
}\\[8mm]
P.~P.~Novichkov\(^{\,a,}\)\footnote{E-mail: \texttt{pavel.novichkov@ipht.fr}},
J.~T.~Penedo\(^{\,b,}\)\footnote{E-mail: \texttt{joao.t.n.penedo@tecnico.ulisboa.pt}},
S.~T.~Petcov\(^{\,c,d,}\)\footnote{Also at
Institute of Nuclear Research and Nuclear Energy,
Bulgarian Academy of Sciences, 1784 Sofia, Bulgaria.}\\
 \vspace{5mm}
\(^{a}\)\,{\it Institut de Physique Théorique, CEA, CNRS, Université Paris--Saclay,\\
F--91191 Gif-sur-Yvette cedex, France} \\
\vspace{2mm}
\(^{b}\)\,{\it CFTP, Departamento de Física, Instituto Superior Técnico, Universidade de Lisboa,\\
Avenida Rovisco Pais 1, 1049-001 Lisboa, Portugal} \\
\vspace{2mm}
\(^{c}\)\,{\it SISSA/INFN, Via Bonomea 265, 34136 Trieste, Italy} \\
\vspace{2mm}
\(^{d}\)\,{\it Kavli IPMU (WPI), UTIAS, The University of Tokyo, \\
Kashiwa, Chiba 277-8583, Japan}
\end{center}
\vspace{2mm}

\begin{abstract}
We study the problem of modulus stabilisation in the framework of the modular symmetry approach to the flavour problem. 
By analysing simple UV-motivated CP-invariant potentials 
for the modulus \(\tau\)
we find that a class of these potentials has (non-fine-tuned) CP-breaking minima in the vicinity of the point of 
\(\mathbb{Z}_3^{ST}\) residual symmetry, \(\tau \simeq e^{2\pi i/3}\).
Stabilising the modulus at these novel minima breaks spontaneously the CP symmetry and can naturally explain the mass hierarchies of charged leptons and possibly of quarks.

\end{abstract}

\end{titlepage}
\setcounter{footnote}{0}
%

\section{Introduction}
\label{sec:intro}

Understanding the origins of flavour in both the
quark and lepton sectors, i.e.,~of the patterns
of quark masses and mixing, of charged-lepton 
and neutrino masses and of neutrino mixing and  
CP violation in the quark and lepton sectors, 
is one of the most challenging fundamental problems
in contemporary particle physics~\cite{Feruglio:2015jfa}.
Although this problem arose more than 20 years ago, 
all efforts to find a satisfactory solution
have essentially failed.
A universal, elegant, natural and viable theory of flavour 
is still lacking. Constructing such a theory would be a major 
achievement and a breakthrough in particle physics.

The unsatisfactory status of the flavour
problem together with the remarkable progress 
made in the studies of neutrino oscillations
(see, e.g.,~\cite{PDG2019}),
which began 23 years ago with the discovery 
of oscillations of atmospheric \(\nu_{\mu}\) and 
\(\bar{\nu}_{\mu}\) by the SuperKamiokande
experiment~\cite{Fukuda:1998mi} and led
to the determination of the pattern of 3-neutrino mixing
consisting of two large and one small mixing angles,
stimulated renewed attempts to seek  
new approaches to the lepton 
as well as to the quark flavour problems.
A step forward in this direction was made in 2017
in Ref.~\cite{Feruglio:2017spp}, where
the idea of using modular invariance as a flavour 
symmetry was put forward.
The first phenomenologically viable lepton flavour models 
based on modular symmetry appeared in the first half of 2018~\cite{Kobayashi:2018vbk,Penedo:2018nmg,Criado:2018thu}
and, since then, the modular-invariance approach 
to the flavour problem has been and continues to be 
intensively investigated and developed with encouraging results. 

In the modular-invariance approach, 
the elements of the Yukawa coupling 
and fermion mass matrices in the Lagrangian
are expressed in terms of modular forms of a certain level \(N\) 
and a limited number of coupling constants.
The modular forms are functions of a single complex scalar field \(\tau\)
--- the modulus. Both the modulus \(\tau\) and the modular forms  
have specific transformation properties under the action of 
the modular group \(\Gamma \equiv SL(2,\mathbb{Z})\).
The matter fields are assumed to transform in representations 
of an inhomogeneous (homogeneous) 
finite modular group \(\Gamma^{(\prime)}_N\), 
while the modular forms furnish 
irreducible representations of the same group.
For \(N\leq 5\), the finite modular groups \(\Gamma_N\) 
are isomorphic to the permutation groups 
\( S_3\), \( A_4\), \( S_4\) and \( A_5\) 
(see, e.g.,~\cite{deAdelhartToorop:2011re}) 
and the groups \(\Gamma^\prime_N\) are isomorphic to
their double covers.
These groups are quite extensively used in flavour model building 
(see, e.g.,~\cite{Altarelli:2010gt,Ishimori:2010au,King:2014nza,Tanimoto:2015nfa,Petcov:2017ggy}).
The modular symmetry described 
by the finite modular group \(\Gamma^{(\prime)}_N\) plays 
the role of a flavour symmetry and 
the theory is assumed to be invariant under 
the whole modular group \(\Gamma\).

A very appealing feature of this approach is that   
the vacuum expectation value (VEV) of the modulus \(\tau\) 
can be the only source of flavour
symmetry breaking, such that flavons are not needed.%
\footnote{The first modular-invariant ``minimal'' lepton flavour model without flavons was constructed in~\cite{Penedo:2018nmg}.
}
Another appealing feature of the discussed framework 
is that the VEV of \(\tau\) can also be the only source 
of breaking of  CP symmetry~\cite{Novichkov:2019sqv}.

There is no VEV of \(\tau\) which preserves the full modular symmetry.
However, as pointed out in~\cite{Novichkov:2018ovf}
and exploited 
in~\cite{Novichkov:2018yse,Novichkov:2018nkm,Okada:2020brs},
there exist three values in the modular group 
fundamental domain, which do not break the 
modular symmetry completely.
These so-called ``fixed points'' are
\(\tau_\text{sym} = i,\, \omega, \,i \infty\),
with \(\omega \equiv \exp(2\pi i / 3)
= -1/2 + \sqrt{3}/2\,i\) (the `left cusp'), and, 
for theories based on \(\Gamma_N\) invariance,
preserve \(\mathbb{Z}^{S}_2\),
\(\mathbb{Z}^{ST}_3\), and 
\(\mathbb{Z}^{T}_N\) residual symmetries, respectively.%
\footnote{ 
In the case of the double cover groups \(\Gamma^\prime_N\),
these residual symmetries are augmented by
\(\mathbb{Z}_2^R\)~\cite{Novichkov:2020eep}.
}
After the flavour symmetry is fully or partially broken, 
the modular forms and thus 
the elements of the Yukawa coupling
and fermion mass matrices get fixed. 
Correspondingly, the fermion mass matrices 
exhibit a certain symmetry-constrained
flavour structure.

The approach to the flavour problem based on 
modular invariance has been widely explored
so far primarily in the framework 
of supersymmetric (SUSY) theories.
Within the framework of rigid (\(\mathcal{N}=1\)) SUSY,
modular invariance is 
assumed to be a property of 
the superpotential, whose holomorphicity
restricts the number of allowed terms.
Following a bottom-up approach,  
phenomenologically viable and ``minimal'' 
lepton flavour models based on modular symmetry, 
which do not include flavons,  
have been constructed first using the groups 
\(\Gamma_4 \simeq S_4\)~\cite{Penedo:2018nmg} and 
\(\Gamma_3 \simeq A_4\)~\cite{Criado:2018thu}.
A ``non-minimal'' model with flavons based on 
\(\Gamma_2 \simeq S_3\)
has been proposed in~\cite{Kobayashi:2018vbk}.
After these studies, the interest in the approach 
grew significantly and a large variety of models has
been constructed and extensively studied.
This includes:%
\footnote{A rather complete list of the articles on modular-invariant 
models of lepton and/or quark flavour, which appeared by March of 2021, 
can be found in~\cite{Novichkov:2021evw}.
We cite here only a representative sample.
}
\begin{enumerate}[i)]
    \item lepton flavour models based on the groups 
\(\Gamma_4 \simeq S_4\)~\cite{Novichkov:2018ovf, Kobayashi:2019mna,Kobayashi:2019xvz,Gui-JunDing:2019wap},
\(\Gamma_5 \simeq A_5\)~\cite{Novichkov:2018nkm, Ding:2019xna},
\(\Gamma_3 \simeq A_4\)~\cite{Kobayashi:2018scp, Novichkov:2018yse, Ding:2019zxk, Gui-JunDing:2019wap, Kobayashi:2019gtp},
\(\Gamma_2 \simeq S_3\)~\cite{Okada:2019xqk}
and \(\Gamma_7 \simeq PSL(2,\mathbb{Z}_7)\)~\cite{Ding:2020msi},
\item models of quark flavour~\cite{Okada:2018yrn} and of quark--lepton 
unification~\cite{Kobayashi:2018wkl,Okada:2019uoy,Kobayashi:2019rzp,Lu:2019vgm,Okada:2020rjb,Chen:2021zty},
\item models with multiple moduli, considered first phenomenologically
in~\cite{Novichkov:2018ovf, Novichkov:2018yse} and further studied, e.g.,
in~\cite{deMedeirosVarzielas:2019cyj,King:2019vhv,Ding:2020zxw},
\item models in which the formalism of the interplay of modular and 
generalised CP (gCP) symmetries, developed and applied first to the 
lepton flavour problem in~\cite{Novichkov:2019sqv},
is explored~\cite{Kobayashi:2019uyt,Okada:2020brs,Yao:2020qyy,Wang:2021mkw,Ding:2021iqp}.
\end{enumerate}
Also the formalism of the double cover finite modular 
groups~\(\Gamma'_N\), to which top-down constructions typically lead
(see, e.g.,~\cite{Nilles:2020nnc,Kikuchi:2020nxn} and references therein),
has been developed and viable flavour models have been constructed
for the cases of \(\Gamma'_3 \simeq T'\)~\cite{Liu:2019khw},
\(\Gamma'_4 \simeq S'_4\)~\cite{Novichkov:2020eep,Liu:2020akv} and 
\(\Gamma'_5 \simeq A'_5\)~\cite{Wang:2020lxk,Yao:2020zml}.
Recently, the framework has been further generalised to arbitrary finite modular groups (i.e.,~those not described by series \(\Gamma_N^{(\prime)}\)) in Ref.~\cite{Liu:2021gwa}.
It is hoped that the results obtained in the bottom-up 
modular-invariant approach to the lepton and quark flavour
problems will eventually connect with top-down results
(see, e.g.,~\cite{Kobayashi:2018rad, Kobayashi:2020hoc, Abe:2020vmv, Ohki:2020bpo, Nilles:2020kgo, Nilles:2020tdp,Ishiguro:2020nuf,Ishiguro:2020tmo, Baur:2020yjl, Almumin:2021fbk,Baur:2021bly}),
based on UV-complete theories.

\vskip 2mm

In practically all phenomenologically viable 
lepton and/or quark flavour models based on modular invariance,
the VEV of the modulus is treated as a free parameter 
which is determined by confronting model predictions with 
experimental data.
Its value is critical for phenomenological viability
and can vary significantly, depending on the model.
For instance, viable \(\Gamma_4 \simeq S_4\) lepton flavour 
models consistent with 
the available data on lepton masses and mixing
have been obtained in~\cite{Novichkov:2018ovf}
for values of \(\tau\)
relatively close to the symmetric point 
\(\tau_\text{sym} = i\), 
very close to the boundary of the fundamental domain,
at \(\re\tau \simeq \pm 0.5\), 
as well as for \(\tau \simeq \pm 0.143 + 1.523\, i \) 
and  \(\tau \simeq \pm 0.179 + 1.397\,i\). 
In~\cite{Yao:2020qyy}, where modular \(A_4\) lepton and quark 
flavour models have been considered, the authors find 
viable models for different values of \(\tau\) close to 
\(\tau_\text{sym} = i\), and values of \(\tau\)
close to the imaginary axis (\(\re\tau=0\))
with rather large \(\im\tau \simeq 2.67\). 
In~\cite{Okada:2020ukr}, viable lepton flavour models with \(A_4\)
modular symmetry have been presented for 
values of \(\tau\) close to each of the three 
symmetric points, \(\tau_\text{sym} = i, \omega, i\infty\).
It should be clear from this discussion that determining 
the VEV of \(\tau\) from first principles and not 
from fits to the data could be used as a powerful  
selection criteria for the proposed flavour models.

An additional unique feature of the modular approach 
to the flavour problem,
as shown recently in Ref.~\cite{Novichkov:2021evw}, is
that one can obtain fermion (charged-lepton and quark) 
mass hierarchies from the properties of the modular forms ---
without fine-tuned constants ---
provided the VEV of the modulus \(\tau\) 
takes a value close to the one of the symmetric points 
\(\tau_\text{sym} = \omega\) (the left cusp)
or \(\tau_\text{sym} = i\infty\).%
\footnote{
Given the exponential dependence of modular forms on \(\im\tau\),
``\(\infty\)'' effectively means a number 
sufficiently bigger than one, e.g., a number \(\sim (2-3)\).
}
In practically all modular-invariant
flavour models without flavons considered before 
the appearance of~\cite{Novichkov:2021evw},  
the charged-lepton and quark mass hierarchies 
were successfully reproduced with the help of severe 
fine-tuning of the limited number of coupling constants 
present in the Yukawa couplings.
In the viable fine-tuning-free model 
constructed in~\cite{Novichkov:2021evw}, 
data requires \(\tau\) to have a value near the cusp 
\(\tau_\text{sym} = \omega\), selecting a best-fit point with
\begin{equation}
\tau \simeq -\,0.496 + 0.877\,i\,.
\label{eq:FTFvev}
\end{equation}
%
The viable region for \(\tau\) actually corresponds
to a small ring around the cusp, of radius \(|u| \simeq 0.007\),
with \(u \equiv (\tau-\omega)/(\tau-\omega^2)\).
The smallness of this quantity is at the basis
of the mechanism giving rise to fermion mass hierarchies
in this context.
The question to address is whether such data-driven
values of \(\tau\) can be naturally justified
by a dynamical principle, e.g., from a top-down perspective.
\vskip 2mm

Attempts to determine the value of \(\tau\)
on the basis of dynamical considerations  
were made in, e.g.,~\cite{Kobayashi:2019xvz,Kobayashi:2019uyt,Ishiguro:2020tmo}.
In~\cite{Kobayashi:2019xvz}, the authors consider lepton flavour 
models with \(\Gamma_3\simeq A_4\) modular symmetry
arising from the breaking of \(\Gamma_4\simeq S_4\) 
symmetry by anomalies. Fitting the available data
on charged-lepton masses, neutrino mixing angles
and neutrino mass-squared differences, they determine
the values of the modulus for which the models are
phenomenologically viable.   
They further attempt to obtain these values 
within the supergravity framework,
constructing relatively simple superpotentials.
The latter are assumed to be generated non-perturbatively 
by hidden sector dynamics and involve singlet modular forms
of weights 4 or 6. Only the linear combination of potentials,
each involving one of the two singlet modular forms
is shown to have absolute minima at some of the requisite
CP-nonconserving values of \(\tau\).
However, this combination effectively contains one
additional complex parameter which violates CP 
symmetry explicitly.
In a follow-up study~\cite{Kobayashi:2019uyt}, 
the possibility of spontaneously breaking the CP symmetry in
theories of flavour based on \(\Gamma_3 \simeq A_4\), 
\(\Gamma_2 \simeq S_3\) and \(\Gamma_4 \simeq S_4\) 
was analysed.
Superpotentials analogous to those used in~\cite{Kobayashi:2019xvz}
were constructed, leading, however, to CP-invariant potentials for 
the modulus \(\tau\). The authors of~\cite{Kobayashi:2019uyt}
have found these potentials to have absolute minima at
different CP-conserving values of \(\tau\), 
related by the \(T\) transformation (\(\tau \xrightarrow{T} \tau +1\)).
The same result was found in~\cite{Kobayashi:2019uyt}
to hold also in theories with global supersymmetry and
essentially  the same superpotentials. 
In \cite{Ishiguro:2020tmo} the authors
have considered three-form fluxes in Type IIB string theory and derived
the preferred values of \(\tau\) by investigating
the possible configurations of flux compactifications
on a \(T^6/(Z_2\times Z^\prime_2)\) orbifold 
(exploring the so-called ``string landscape'').
The number of stable vacua depends on a certain
positive integer \(N^{\mathrm{max}}_{\mathrm{flux}}\)
and reads 312, 2918 and 2886221, for 
\(N^{\mathrm{max}}_{\mathrm{flux}}= 10, 100\) and \(1000\), respectively.
These vacua correspond to stabilised values of the modulus \(\tau\)
in the fundamental domain of the modular group. The most probable
of these are found to lie on the border of the fundamental domain
\(\re\tau = -1/2\), on the imaginary axis \(\re\tau = 0\),
and on the arc, \(\tau = \exp(i\alpha)\) 
with, e.g.,~\(\cos\alpha = -\,1/4\).
All these values are CP-conserving~\cite{Novichkov:2019sqv}.
Actually, the so-derived stabilising values of \(\tau\) are shown to cluster 
at the CP-conserving symmetry point \(\tau = \omega\).%
\footnote{See also~\cite{Abe:2020vmv}, where 
\(\tau\) is found to be stabilised at the 
CP-conserving right cusp, \(\tau = \exp(i\pi/3)\).
}

The possibility of spontaneous 
breaking of the CP symmetry by the modulus VEV 
was investigated in~\cite{Bailin:1997fh} 
in a supergravity framework in which the dilaton is also present.
The authors showed that if the dilaton is stabilised by 
non-perturbative corrections to the Kahler potential, 
by varying the four parameters present in 
the relevant effective potential it is possible to find minima of the 
potential at CP-violating VEVs of the modulus \(\tau\) inside 
the fundamental domain of the modular group 
close to the symmetric point \(\tau = i\).

\vskip 5mm

In the present article we address the problem of modulus stabilisation
by analysing known and relatively simple supergravity-motivated 
modular- and CP-invariant potentials for the modulus \(\tau\).
In~\cref{sec:framework} we describe
the framework we employ,
giving some details on how modular symmetry
can act as a flavour symmetry (\cref{sec:modframework}).
We introduce the modular- and CP-invariant potentials
for \(\tau\) in~\cref{sec:scalarpot} and derive
in~\cref{sec:quexpansions} their \(q\)- (and \(u\)-) expansions 
which prove useful for 
the analyses of the potentials. 
We present the main results of our 
study  in~\cref{sec:results},
including the found novel global CP-breaking minima
of the considered potentials.
\Cref{sec:summary} contains a summary of our results.
Some technical details are given in~\cref{app:mod_funcs,app:rigid,app:uexp}.

\vspace{0.5cm}
\section{Framework}
\label{sec:framework}

\subsection{Modular symmetry as a flavour symmetry}
\label{sec:modframework}

The modular approach to flavour is based on invariance under the action of the modular group \(\Gamma \equiv SL(2, \mathbb{Z})\). While in this \namecref{sec:modframework} we summarise the defining features of this (bottom-up) framework, the reader is referred to~\cite{Novichkov:2021evw, Novichkov:2020eep} for a more detailed description.

The modular group  is generated by three elements \(S\), \(T\) and \(R\) obeying \((ST)^3 = R^2 =\mathbb{1}\), \(S^2 = R\) and \(RT = TR\).
A generic element \(\gamma\) of this group acts on the modulus chiral superfield \(\tau\) as a fractional linear transformation,
\begin{equation}
  \gamma = \begin{pmatrix} a & b \\ c & d \end{pmatrix} \in \Gamma: \qquad
  \tau \to  \gamma \tau = \frac{a\tau + b}{c\tau + d} \,.
  \label{eq:gamma}
\end{equation}
Its action on matter superfields instead reads~\cite{Ferrara:1989bc,Ferrara:1989qb,Feruglio:2017spp}
\begin{equation}
  \gamma\in \Gamma: \quad\psi_i \to (c\tau + d)^{-k} \, \rho_{ij}(\gamma) \, \psi_j \,,
\end{equation}
where \(k\) is the modular weight of \(\psi\) and \(\rho\) is a unitary representation of \(\Gamma\). Modular symmetry plays the role of a discrete flavour symmetry 
when \(\rho(\gamma)=\mathbb{1}\) for \(\gamma \equiv \mathbb{1}\,(\text{mod}\,N)\).
In this case \(\rho\) is effectively a (unitary) representation of the finite quotient group \(\Gamma'_N \simeq SL(2, \mathbb{Z}_N)\) characterised by the integer level \(N\geq 2\). 
For \(N\leq 5\), \(\Gamma^\prime_{2,3,4,5}\) are isomorphic to the double covers \(S^\prime_3, A^\prime_4, S^\prime_4, A^\prime_5\) of the well-known permutation groups \(S_3, A_4,S_4,A_5\), 
with \(S^\prime_3\equiv S_3\).
If the \(R\) generator acts trivially on fields, one is instead dealing with the inhomogeneous finite modular group \(\Gamma_N\), a quotient of 
\(\overline{\Gamma} \simeq PSL(2,\mathbb{Z}) 
\simeq SL(2,\mathbb{Z})/\mathbb{Z}^R_2\).
For small \(N\), the latter finite groups are isomorphic to the 
already quoted permutation groups \(\Gamma_{2,3,4,5} \simeq S_3, A_4, S_4, A_5\). 

Modular symmetry may determine the structure of fermion mass matrices, as it severely constrains the form of the superpotential \(W(\tau,\psi)\), thanks also to 
its holomorphicity.
To ensure the modular transformation properties of \(W\), Yukawa couplings and fermion mass terms must generically depend on \(\tau\) and transform in a way similar to fields --- they are multiplets \(Y(\tau)\) of modular forms of level \(N\), characterised by their own 
integer weights \(k_Y > 0\)
and representations 
of the flavour symmetry group \(\Gamma^{(\prime)}_N\).
Since the number of available independent modular forms is finite (and small for small \(k_Y\)), only a limited set may contribute to \(W\). Thus, the number of superpotential parameters is restricted, and mass and coupling matrices are determined once the lowest (scalar) component of the modulus \(\tau\) acquires a VEV. This setup can be remarkably predictive.%
\footnote{
One needs to control the form of the (joint) Kähler potential of the modulus and matter fields \(K(\tau,\psi)\)~\cite{Chen:2019ewa},
typically taken to have a minimal form in a bottom-up approach.
This problem is the subject of ongoing research (see, e.g.,~\cite{Nilles:2020nnc,Chen:2021prl}).
}

\begin{figure}[t]
  \centering
  \includegraphics[width=0.6\textwidth]{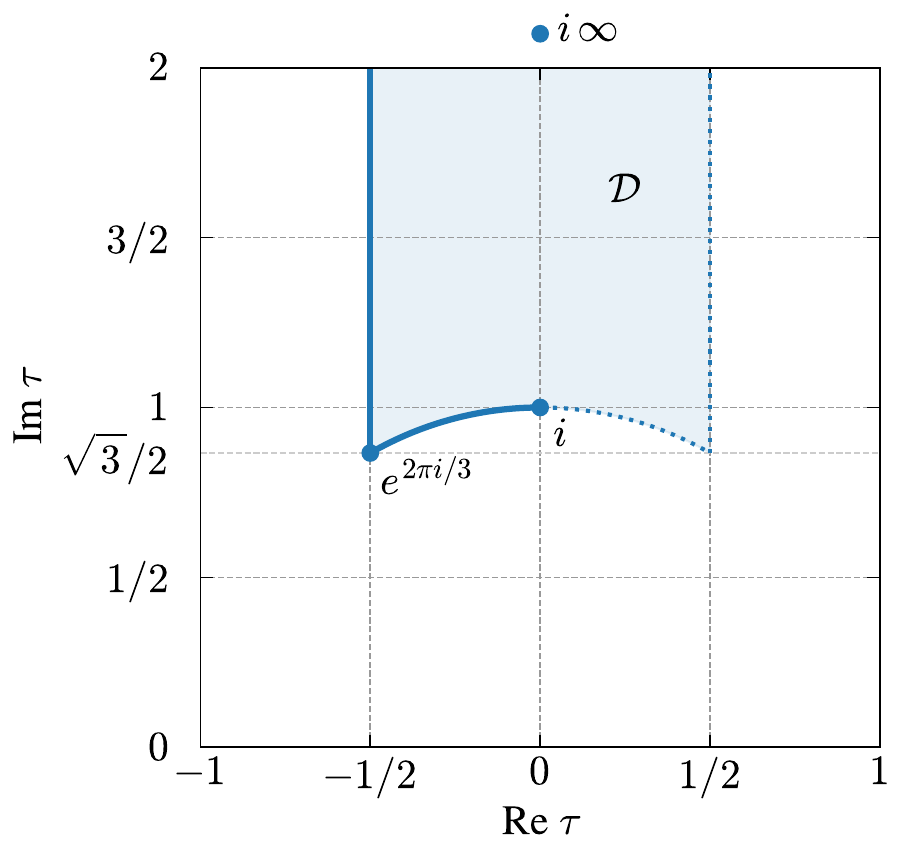}
  \caption{The fundamental domain \(\mathcal{D}\) of the modular group \(\Gamma\) and its three symmetric points \(\tau_\text{sym} = i\, \infty, i, \omega\). The value of \(\tau\) can be restricted to \(\mathcal{D}\) by a suitable modular transformation. 
  (Figure from Ref.~\cite{Novichkov:2020eep}.)
}
  \label{fig:fund_domain}
\end{figure}

The breakdown of modular symmetry is parameterised by the VEV of the modulus (\(\im \tau >0\)).%
\footnote{Flavon fields are not required in this approach, and we do not consider them here.}
This VEV can always be restricted to the fundamental domain~\(\mathcal{D}\) of the modular group,
defined by the union
\begin{equation}
\label{eq:fund_domain}
\mathcal{D} \equiv \left\{ \tau \in \mathcal{H} : -\frac{1}{2} \leq \re \tau < \frac{1}{2},\, |\tau| > 1 \right\} \cup \left\{ \tau \in \mathcal{H} : -\frac{1}{2} < \re \tau \leq 0,\, |\tau| = 1 \right\} \,,
\end{equation}
see also~\cref{fig:fund_domain}. 
%
%
Any choice of the VEV of the modulus in the upper-half plane can be related to a single \(\tau \in\mathcal{D}\) via a modular transformation, and it is therefore physically equivalent to it (see also \crefext{section}{4} of~\cite{Novichkov:2018ovf}). 
Instead, two elements in the fundamental domain cannot be related by a modular transformation and are thus physically inequivalent. By convention, the right half of the boundary of \(\mathcal{D}\) --- including the right half of the unit arc ---
is excluded from the above definition, since it is equivalent to the left half.%
\footnote{
 Any point on the right boundary \(\re\tau = 1/2\) can be obtained 
 from a point with the same \(\im\tau\) on the left boundary 
 \(\re\tau = -\,1/2\) by a \(T\)-transformation, while any point on 
 the right half of the arc with given \(\re\tau > 0\) and \(\im\tau\) 
 can be obtained by an \(S\)-transformation from the point 
 on the left half of the arc with the same \(\im\tau\).
 }

Even though there is no value of \(\tau\) which preserves the full modular symmetry, 
as we have already noted,
specific residual symmetries remain 
at certain symmetric points \(\tau = \tau_\text{sym}\).
There are only three (inequivalent) symmetric points~\cite{Novichkov:2018ovf}:%
\footnote{The \(R\) generator is unbroken for any value of \(\tau\) and a \(\mathbb{Z}_2^R\) symmetry is always preserved~\cite{Novichkov:2020eep}.}
\begin{itemize}
\item \(\tau_\text{sym} = i\), invariant under \(S\), preserving \(\mathbb{Z}_4^S \) (note that \(S^2 = R\)); 
\item \(\tau_\text{sym} = i \infty\), invariant under \(T\), preserving \(\mathbb{Z}_N^T \times \mathbb{Z}_2^R\); and
\item \(\tau_\text{sym} = \omega = \exp (2\pi i / 3)\) (the ``left cusp''), invariant under \(ST\), preserving \(\mathbb{Z}_3^{ST} \times \mathbb{Z}_2^R\).
\end{itemize}
In models where \(\tau\) deviates slightly from one of these values \(\tau_\text{sym}\),
fermion mass hierarchies may be generated as powers of the small deviation \(\lvert \tau-\tau_\text{sym} \rvert\) (or as powers of \(|q|=e^{-2\pi \im\tau}\) in the case \(\tau_\text{sym}=i\infty\))~\cite{Novichkov:2021evw}.%
\footnote{The residual symmetries are at play at the level of the whole action, and corrections to the Kähler are not expected to qualitatively affect the hierarchies in the fermion mass spectrum.}
Finally, these symmetric values preserve the CP symmetry of
a CP- and modular-invariant theory~\cite{Novichkov:2019sqv,Novichkov:2020eep}. In such a theory, the \(\mathbb{Z}^\text{CP}_2\) symmetry is preserved for \(\re\tau = 0\) or for \(\tau\) lying on the border of \(\mathcal{D}\), but is broken for other generic values of the modulus.

\subsection{Scalar potential}
\label{sec:scalarpot}

While in bottom-up approaches the VEV of \(\tau\) is scanned over the fundamental domain in a fit to the data, it is clearly desirable to have a dynamical reason for its specific, phenomenologically viable value(s).
This is the issue we address here.
In this work we analyse a known class of simple modular-invariant potentials 
which are functions of \(\tau\) alone. Since we are concerned only with the contribution involving the modulus, these turn out to be simplified models, which nevertheless are explicit examples of \(\mathcal{N} = 1\) supergravity models.%
\footnote{
A fully realistic string compactification is expected to involve other moduli, as well as gauge bosons and matter fields~\cite{Gonzalo:2018guu}.
Therefore, the investigated potentials correspond to a subsector of the full theory.
In particular, we do not identify their minimum values with the cosmological constant, as the latter receives contributions from other subsectors as well.
}
We thus focus on the simple Kähler potential~\cite{Cvetic:1991qm},
    \begin{equation}
    K(\tau,\overline{\tau}) = - \Lambda_K^2 \log(2\, \im\tau)\,,
        \label{eq:Kahler}
    \end{equation}
where \(\Lambda_K\) is a scale (mass dimension one).

We discuss next the form of simple superpotentials \(W(\tau)\), following Refs.~\cite{Cvetic:1991qm,Gonzalo:2018guu}. Keeping in mind 
the single-modulus case, the relevant \(\mathcal{N} = 1\) supergravity action depends on the Kähler-invariant function
\begin{align}
    G(\tau,\overline{\tau}) \,=\, 
    \kappa^2 K(\tau,\overline{\tau}) + 
    \log \left\lvert \kappa^3 W(\tau) \right\rvert^2\,,
\end{align}
where \(\kappa^2 = 8\pi/M_P^2\), \(M_P\) being the Planck mass.
Given the choice of~\cref{eq:Kahler} for the Kähler potential, modular invariance of \(G\) implies that the superpotential \(W\) carries modular weight \(-\mathfrak{n}\), where \(\mathfrak{n}=\kappa^2 \Lambda_K^2\). We consider integer values of \(\mathfrak{n}\), in line with~\cite{Cvetic:1991qm,Gonzalo:2018guu}. The superpotential can then be parameterised in terms of the Dedekind \(\eta\) function (see~\cref{app:mod_funcs}) and a modular-invariant function \(H\), as
\begin{equation}
W(\tau) \,=\, \Lambda_W^3 \,\frac{H(\tau)}{ \eta(\tau)^{2\mathfrak{n}}}\,,
\label{eq:superpotential}
\end{equation}
where \(\Lambda_W\) is a mass scale so that \(H(\tau)\) is dimensionless.
The most general \(H\) (without singularities in the fundamental domain) can be cast in the following form~\cite{Cvetic:1991qm}:
\begin{align}
H(\tau)  
\,=\, \left ( j(\tau)-1728 \right )^{m/2}j(\tau)^{n/3}\, \mathcal{P}\left( j(\tau) \right )\,,
\label{eq:H}
\end{align}
making use of the Klein \(j\) function, which is invariant under the action of the modular group \(SL(2,\mathbb{Z})\) (see~\cref{app:mod_funcs}). Here, \(m\) and \(n\) are non-negative integers and \(\mathcal{P}\) is a polynomial in \(j(\tau)\).
    
The scalar potential in \(\mathcal{N} = 1\) supergravity is given by (see, e.g.,~\cite{Ibanez:2012zz})
\begin{equation}
  V = e^{\kappa^2 K} \left( K^{i \, \bar{j}} D_i W D_{\,\bar{j}} W^{*} - 3 \kappa^2 \lvert W \rvert^2 \right) \,,
\label{eq:V1}
\end{equation}
where \(D_i \equiv \partial_i + \kappa^2 (\partial_i K)\), \(K^{i \, \bar{j}}\) is the inverse of the Kähler metric \(K_{i \, \bar{j}} \equiv \partial_i \, \partial_{\,\bar{j}} K\), and \(\partial_i\) (\(\partial_{\,\bar{j}}\)) is the derivative with respect to the corresponding field (its conjugate).
In our setup, the only field is the modulus \(\tau\), and the scalar potential follows from the explicit forms of \(K(\tau,\overline{\tau})\) and \(W(\tau)\) given above,
\begin{align}
    V(\tau, \overline{\tau}) = \frac{\Lambda_V^4}{(2\im\tau)^\mathfrak{n} \lvert \eta(\tau) \rvert^{4\mathfrak{n}}}
    \left[
    \left\lvert
    i H'(\tau) + \frac{\mathfrak{n}}{2\pi} H(\tau) \hat{G}_2(\tau,\overline{\tau})
    \right\rvert^2 \frac{(2 \im\tau)^2}{\mathfrak{n}} - 3\lvert H(\tau) \rvert^2
    \right]\,,
    \label{eq:V2}
\end{align}
where we have defined \(\Lambda_V = \left( \kappa^2 \Lambda_W^6 \right)^{1/4}\) as the mass scale of the potential, and \(\hat{G}_2\) is the non-holomorphic Eisenstein function of weight 2 (see, e.g.,~\cite{Cvetic:1991qm}).
It is given by 
\begin{align}
\hat{G}_2(\tau,\overline{\tau}) = {G}_2(\tau) - \frac{\pi}{\im\tau}\,,
\label{eq:G2hat}
\end{align}
where \(G_2\) is its holomorphic counterpart (see~\cref{app:mod_funcs}).
\(G_2\) can be related to the Dedekind function via
\begin{equation}
    \frac{\eta'(\tau)}{\eta(\tau)} = \frac{i}{4\pi} G_2(\tau)\,.
    \label{eq:G2}
\end{equation}
It is not difficult to show that the potential \(V(\tau, \bar{\tau})\) is modular-invariant.
We briefly discuss its global SUSY limit 
and its minima in this limit in~\cref{app:rigid}.

We are interested in the simple cases investigated in Refs.~\cite{Cvetic:1991qm,Gonzalo:2018guu}, for which \(\mathfrak{n} = 3\) corresponds to the number of compactified complex dimensions.%
\footnote{
The compactification of 6 dimensions may bring about three moduli \(\tau_i\) (\(i=1,2,3\)),
 corresponding to the radii of three two-tori.
For simple potentials symmetric under the exchange of the \(\tau_i\),
the preferred minimum is found to occur at \(\tau_1 = \tau_2 = \tau_3 = \tau\)~\cite{Cvetic:1991qm}. This result gives support to studying the case of only one modulus, as is done here.}
With this choice, the scalar potential reads:
\begin{align}
    V(\tau, \overline{\tau}) =  \frac{\Lambda_V^4}{8(\im\tau)^3 \lvert \eta \rvert^{12}}
    \left[
    \frac{4}{3} \left\lvert
    i H' + \frac{3}{2\pi} H \hat{G}_2
    \right\rvert^2 (\im\tau)^2 - 3 \lvert H \rvert^2
    \right]\,.
    \label{eq:V}
\end{align}
In what follows, we analyse the global minima of this potential. We consider the form of \(H(\tau)\) given in~\cref{eq:H} for different values of \(m\) and \(n\). 
Following again~\cite{Cvetic:1991qm} (see also~\cite{Gonzalo:2018guu}), we take the simplest choice \(\mathcal{P}(j)=1\), which nevertheless yields non-trivial results. 
In this case, the potential \(V\) can be shown to be CP-symmetric,
i.e.,~to be invariant under a reflection with respect to the imaginary axis~\cite{Novichkov:2019sqv}, \(\tau \to - \overline{\tau}\).
This follows from the fact that 
under the reflection \(\tau \to - \overline{\tau}\),
the functions \(H\), \(\eta\), \(j\) and \(\hat{G_2}\)
are transformed to their conjugates
(while \(H' \to - H'^*, \eta'\to -\eta'^*\)). 
This \(\mathbb{Z}_2^\text{CP}\) symmetry is present for a more general choice of \(\mathcal{P}(j)\), provided the polynomial coefficients are real or share a common complex phase.

Despite the modular- and CP- invariance of \(V\), the vacuum,  which  breaks the modular symmetry for any value of the modulus \(\tau\), may also spontaneously break the CP symmetry.
Extrema not lying at CP-conserving points make up an inequivalent (degenerate) pair, 
at some \(\tau\) and \(-\overline{\tau}\). 
In Ref.~\cite{Cvetic:1991qm}, it was conjectured that all extrema of \(V\) would lie at CP-conserving values of \(\tau\), i.e.,~either on the boundary of the fundamental domain \(\mathcal{D}\) or on the imaginary axis. Therein, the cases \((m,n) = (0,0), (1,1), (0,3)\) were explicitly examined. The global minima of the corresponding potentials were indeed found to lie at \(\tau \simeq 1.2\, i\) (imaginary axis), \(\tau \simeq \pm 0.24+0.97\,i\) (equivalent minima on the unit arc) and \(\tau = i\), respectively.
While we have verified these particular results, we have further found that
potentials with \(n=0\) but \(m>0\) do allow for CP-breaking global minima. 
Moreover, these minima are found to be located in the vicinity of the left cusp \(\tau = \omega\), at values of \(\lvert \tau-\omega \rvert\) favoured by the mechanism put forward in~\cite{Novichkov:2021evw} to explain fermion (charged-lepton and quark) mass hierarchies, as we will see in the following sections.

\subsection{\texorpdfstring{\(q\)}{q}- and \texorpdfstring{\(u\)}{u}- expansions}
\label{sec:quexpansions}
%
As a first step in the analysis of the potential given in~\cref{eq:V}, we express the functions \(j\)  (and therefore \(H\), \(H'\)) and \(\hat{G}_2\)  in terms of the Dedekind \(\eta\) and its derivatives. 
Rewriting the latter is immediate via~\cref{eq:G2}, while the former can be expressed as
\begin{align}
  j &= \left[
  \frac{72}{\pi^2 \eta^6} \left( \frac{\eta'}{\eta^3} \right)' \,
  \right]^3 \,,
  \label{eq:jmain}
\end{align}
%
see~\cref{app:mod_funcs}. 
For a broad numerical analysis, it then suffices to know the \(q\)-expansion of \(\eta\), i.e.,~its expansion in powers of \(q= e^{2\pi i \tau}\), up to a certain order. This expansion has the well-known form
\begin{equation}
\begin{aligned}
        \eta
        &= q^{1/24} \sum_{n=-\infty}^{\infty} (-1)^n q^{\frac{3n^2 -n}{2}} 
        = q^{1/24} \left(1 - q - q^{2} + q^{5} + q^{7} - q^{12} - q^{15} + \mathcal{O}(q^{22})\right)\,
\end{aligned}
\label{eq:etaq0}
\end{equation}
and converges rapidly within the fundamental domain, where \(|q|\leq e^{-\sqrt{3}\pi} \simeq 0.004\).

A preliminary \(q\)-expansion analysis of the potential for different
\((m, n)\) with \(n \neq 0\)
reveals CP-conserving global minima on the imaginary axis and on the unit arc, as expected. For \((m,0)\) with \(m\neq0\), it further reveals CP-breaking minima in the vicinity of the left cusp \(\tau=\omega\) (paired with inequivalent minima in the vicinity of the right cusp \(-\overline{\tau} = \omega+1\)). To guarantee a robust numerical analysis as well as an analytical understanding of these CP-breaking minima, we now develop the expansion of the potential in terms of a parameter \(u \equiv (\tau-\omega)/(\tau-\omega^2)\), which quantifies the deviation of \(\tau\) from the left cusp.

This effort is also warranted in light of the results of Ref.~\cite{Novichkov:2021evw}, where the same convenient parameterisation of deviations from the left cusp was motivated and used.
In particular, stabilising \(\tau\) in the vicinity of \(\omega\) or at a point with large \(\im\tau\) can provide a non-fine-tuned, i.e.,~natural explanation of 
the three generation charged-lepton and quark mass hierarchies, based on the smallness of \(|u|\) or \(|q|\).
In contrast, the stabilisation of \(\tau\) in the vicinity of \(i\), even if possible, cannot offer an explanation of these mass hierarchies
in terms of small deviations from this symmetry point
without severe fine-tuning~\cite{Novichkov:2021evw} or some additional non-minimal input~\cite{Feruglio:2021dte}.
Finally, it is known that the potential under analysis diverges for large \(\im\tau\)~\cite{Cvetic:1991qm}, 
leaving \(\tau \simeq \omega\) as the most interesting case to investigate. 

We have seen that the potential can be fully expressed in terms of \(\eta\) and its derivatives. To obtain its \(u\)-expansion, it is enough to determine the \(u\)-expansion of \(\eta\).
Using \(\tau = \omega^2(\omega^2 -u)/(1-u)\), one can write \(\eta\) as a function of \(u\).
It further proves useful to define%
\footnote{Unless explicitly stated, we always take the principal branch of the roots appearing in our discussion.}
\(\tilde\eta(u) = (1-u)^{-1/2} \eta(u)\), since symmetry dictates \(\tilde\eta\) to be a power series in \(u^3\). 
Indeed, it is easy to show using \(\tau \xrightarrow{T} \tau +1\) and \(\tau \xrightarrow{S} -1/\tau\) that
under the modular transformation \(\gamma = ST\) the variable \(u\) transforms as \(u\xrightarrow{ST} \omega^2 u\). Taking 
further into account that 
\(\eta(T\tau) = \exp(i\pi/12)\,\eta(\tau)\) and 
\(\eta(S\tau) = \sqrt{-i\tau}\, \eta(\tau)\), 
one obtains
\(\eta(ST \tau) = \sqrt{-\omega(\tau+1)}\,\eta(\tau)\).
Thus, we have
\begin{equation}
    \tilde\eta(u) \,\xrightarrow{ST}\, \tilde\eta(\omega^2 u)  =  \frac{\sqrt{-\omega(\tau+1)}}{\sqrt{1-\omega^2 u}}\eta(u)
    = \frac{1}{\sqrt{1-u}}\eta(u)  = \tilde\eta(u)\,,
\end{equation}
meaning \(\tilde\eta(u)\) is invariant under \(\gamma=ST\). Given the transformation property of \(u\), it follows that 
only powers of \(u^3\) survive in the \(u\)-expansion of \(\tilde\eta\). Each coefficient in this expansion
is given by its own power series in \(q(\omega)=-e^{-\sqrt{3}\pi}\), which
can be determined by expressing \(q\) in terms of \(u\) in the known \(q\)-expansion for \(\eta\),~\cref{eq:etaq0}.
One can prove analytically that these coefficients are real up to a common phase.
Numerically, we obtain
\begin{equation}
  \label{eq:tildeuexp}
\begin{aligned}
\tilde\eta(u)
&\simeq 
e^{-i \pi  / 24} \left(
0.800579 - 0.573569 u^3 - 0.780766 u^6 - 0.150007 u^9
\right) + \mathcal{O}(u^{12})
\\
&\equiv
e^{-i \pi  / 24} \left(
\tilde\eta_0 + \tilde\eta_3 u^3 + \tilde\eta_6 u^6 + \tilde\eta_9 u^9
\right) + \mathcal{O}(u^{12})
\,,
\end{aligned}
\end{equation}
%
which is an expansion in powers of \(u^3\), as anticipated.
As a final step,
recall that \(\eta(u) = \sqrt{1-u}\, \tilde\eta(u)\), so its \(u\)-expansion is trivially related to that of \(\tilde{\eta}\).
The \(u\)-expansions of other modular forms are collected in~\cref{app:uexp}.

Let us restate the relevance of these results. Knowing the \(u\)-expansion of \(\eta\) allows one to implement the \(u\)-expansion of the potential \(V\).
The use of such an expansion allows for a clear analysis of the shape of \(V\) and of its extrema in the vicinity of the left cusp \(\tau = \omega\), converging faster than the usual \(q\)-expansion. The corresponding results are shown in the next \namecref{sec:results}.

\section{Results}
\label{sec:results}
\subsection{Numerical analysis of minima for various \texorpdfstring{\(m\)}{m}, \texorpdfstring{\(n\)}{n}}
\begin{figure}[tb]
  \centering
  \begin{tabular}{ll}
    \includegraphics[height=0.45\textwidth]{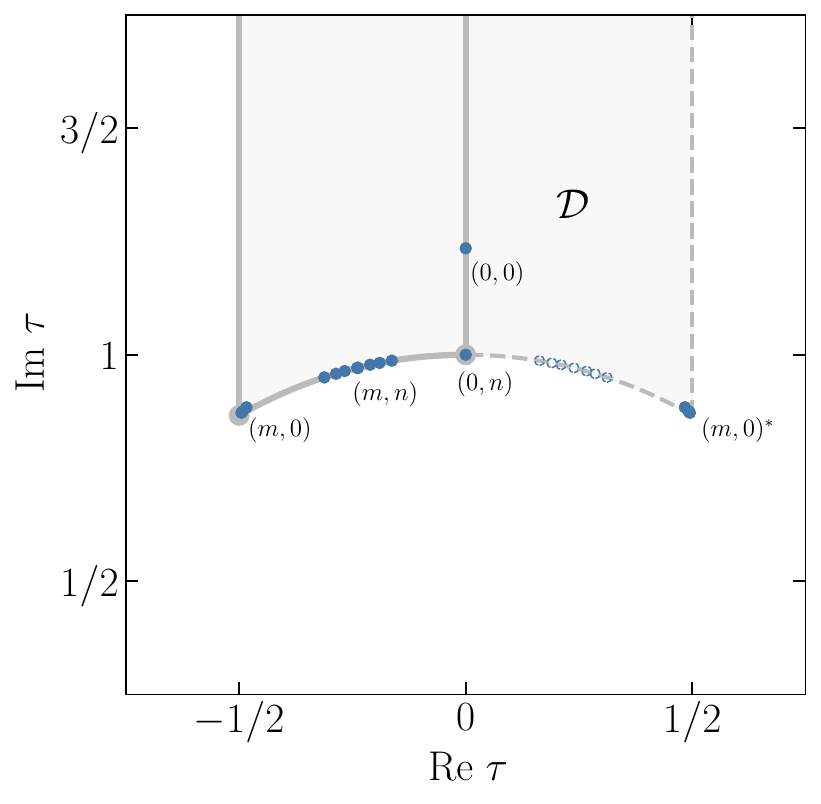} &
    \includegraphics[height=0.45\textwidth]{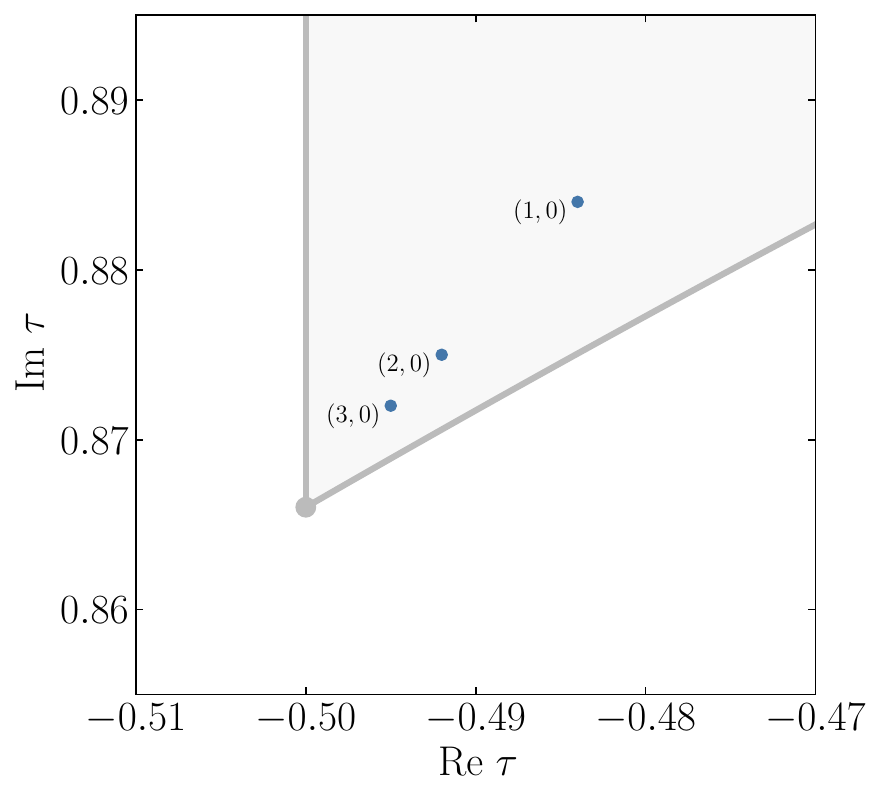}
  \end{tabular}
  \caption{
  Global minima of the potentials \(V(\tau, \bar{\tau})\),~\cref{eq:V}, see text for details.
    Note that points on the right half of the unit arc, which are CP-conjugates of the \((m,n)\) minima, are excluded as they lie outside the fundamental domain.
    The right panel shows the series \((m,0)\) in the vicinity of the left cusp in more detail.}
  \label{fig:summary}
\end{figure}

%
As discussed in~\cref{sec:quexpansions}, the \(q\)-expansions of \(\eta\) and its derivatives allow to compute the potential~\(V(\tau, \bar{\tau})\),~\cref{eq:V}, to arbitrary precision at any point within the fundamental domain~\eqref{eq:fund_domain}.%
\footnote{In practice, we numerically implement a \(q\)-expansion for the potential \(V\). We have checked that the same, stable \(q\)-expansions are obtained independently of expressing \(H\) as a function of \(j\) or as a function of \(\eta\).}
Making use of this fact, we find global minima of the potential numerically for \(0 \leq m, n \leq 3\) (\(\mathcal{P}(j) = 1\)), see~\cref{tab:minima} and~\cref{fig:summary}. As a cross-check, we note that for the special cases of \((m,n) = (0,0), (1,1), (0,3)\) considered in Ref.~\cite{Cvetic:1991qm} our results are consistent with the values reported therein.
This numerical analysis suggests that the minima fall into several classes depending on values of \(m\) and \(n\):
\begin{description}
\item[\(\boldsymbol{(0,0)}\)] is a single minimum at \(\tau \simeq 1.2 i\) on the imaginary axis, corresponding to the case \(m = n = 0\);
\item[\(\boldsymbol{(0,n)}\)] is a single minimum at the symmetric point \(\tau = i\) attained when \(m = 0\), \(n \neq 0\);
\item[\(\boldsymbol{(m,0)}\) and \(\boldsymbol{(m,0)^{*}}\)] are a pair of degenerate minima for each \(m \neq 0\) and \(n = 0\): \((m,0)\) is located in the vicinity of the left cusp \(\tau = \omega\), approaching this symmetric point as \(m\) increases, while \((m,0)^{*}\) is its CP-conjugate;
\item[\(\boldsymbol{(m,n)}\)] is a series of minima on the unit arc, corresponding to \(m \neq 0\), \(n \neq 0\); these minima shift towards \(\tau = \omega\) (\(\tau = i\)) along the arc as \(m\) (\(n\)) grows.
\end{description}
\begin{table}[tb]
  \centering
  \(
  \begin{array}{l@{\qquad}rrrr}
    \toprule
     & \multicolumn{1}{c}{n = 0} & \multicolumn{1}{c}{n = 1} & \multicolumn{1}{c}{n = 2} & \multicolumn{1}{c}{n = 3} \\
    \midrule
    m = 0 & 0.000+1.235 i & 0.000+1.000 i & 0.000+1.000 i & 0.000+1.000 i \\
    m = 1 & \mp 0.484+0.884 i & -0.238+0.971 i & -0.190+0.982 i & -0.163+0.987 i \\
    m = 2 & \mp 0.492+0.875 i & -0.286+0.958 i & -0.239+0.971 i & -0.211+0.978 i \\
    m = 3 & \mp 0.495+0.872 i & -0.312+0.950 i & -0.267+0.964 i & -0.239+0.971 i \\
    \bottomrule
  \end{array}
  \)
  \caption{Values of the modulus \(\tau\) at the global minima of the potential \(V(\tau,\bar{\tau})\),~\cref{eq:V}, obtained numerically for various \(m\) and \(n\).}
  \label{tab:minima}
\end{table}
%
An important observation is that the minima belonging to classes \((0,0)\), \((0,n)\), \((m,n)\) lie either on the boundary of the fundamental domain~\(\mathcal{D}\) or the imaginary axis, in line with the conjecture of Ref.~\cite{Cvetic:1991qm}; these minima are CP-conserving.
However, the \((m,0)^{(*)}\) minima slightly depart from the left (right) cusp symmetric point and the boundary.
This property makes such minima an interesting possibility from the phenomenological viewpoint, as they can naturally explain both CP violation~\cite{Novichkov:2019sqv} and hierarchical mass patterns~\cite{Novichkov:2021evw} in an economical way.
Therefore, we now turn to a discussion of 
the \((m,0)^{(*)}\) minima and the corresponding \(V_{m,0}\) potentials.

%
\subsection{CP-violating minima of \texorpdfstring{\(V_{m,0}\)}{Vm,0}}
\label{sec:CPVminima}
%
Since the \((m,0)^{*}\) minima are trivially related to the \((m,0)\) minima via CP reflection, we concentrate here on the latter series only, i.e.,~we study the behaviour of \(V_{m,0}\) in the vicinity of the left cusp.
The minima of interest deviate only slightly from the boundary; to make sure these deviations are not a numerical artefact of our \(q\)-expansions, we re-expand the potential in terms of \(u\), as described in~\cref{sec:quexpansions}.

Note that \(V_{m,0}\) is a real-valued non-holomorphic function of \(u\), therefore it expands in powers of \(\lvert u \rvert\) rather than \(u\) itself, with coefficients possibly depending on the phase of \(u\).
Denoting this phase as \(\phi\), i.e.,~\(u = \lvert u \rvert e^{i \phi}\), \(\phi \in [-\pi/3, 0]\) (see~\cref{app:uexp}), we find:
\begin{equation}
  \label{eq:Vm0uexp4}
  V_{m,0} = \Lambda_V^4 \, \frac{1728^m}{\sqrt{3} \, \tilde\eta_0^{12}} \,
  \Big\{ {-1} -2 \, \lvert u \rvert^2 + \left( A_m^2 - 3 \right) \lvert u \rvert^4 \Big\} + \mathcal{O}(\lvert u \rvert^6) \,,
\end{equation}
%
where
\begin{equation}
  \label{eq:coefA}
  A_m \equiv \frac{864 \, \lvert \tilde{\eta}_3 \rvert^3}{\pi^6 \, \tilde{\eta}_0^{27}} \, m + \frac{6 \, \lvert \tilde{\eta}_3 \rvert}{\tilde{\eta}_0}
  \simeq 68.78 \, m + 4.30
\end{equation}
%
and \(\tilde{\eta}_i\) are coefficients of the \(u\)-expansion of \(\tilde{\eta}(u)\) defined in~\cref{eq:tildeuexp} (in particular, \(\tilde{\eta}_0 = \lvert \eta(\omega) \rvert\)). 

\begin{figure}[tb]
    \centering
    \includegraphics{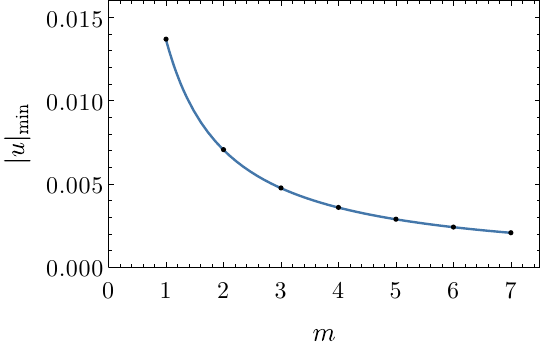}
    \caption{Deviation of the minimum of \(V_{m,0}(\tau, \bar{\tau})\) from the left cusp \(\tau = \omega\) measured by \(\lvert u \rvert = \lvert (\tau - \omega) / (\tau - \omega^2) \rvert\).
    Values obtained numerically (black dots) match the analytical approximation of~\cref{eq:umin} (blue line).}
    \label{fig:umin}
\end{figure}

\begin{figure}[p]
\thisfloatpagestyle{empty}
  \centering
  \begin{tabular}{rl}
    \includegraphics[width=0.42\textwidth]{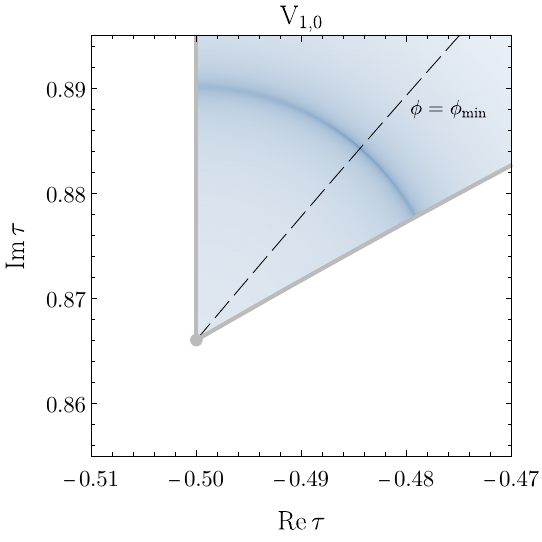} &
    \includegraphics[width=0.42\textwidth]{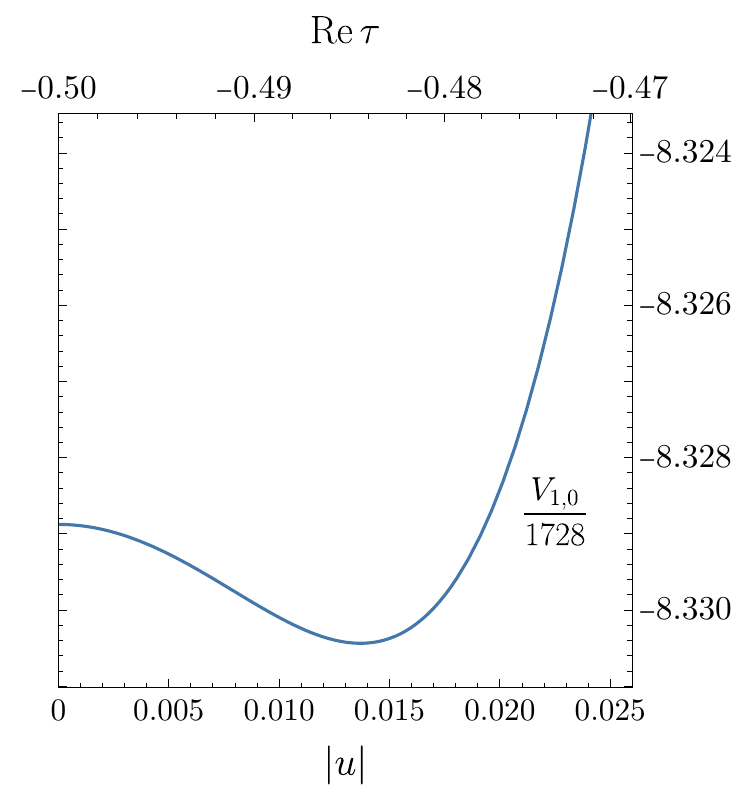}\\[-1ex]
    \includegraphics[width=0.42\textwidth]{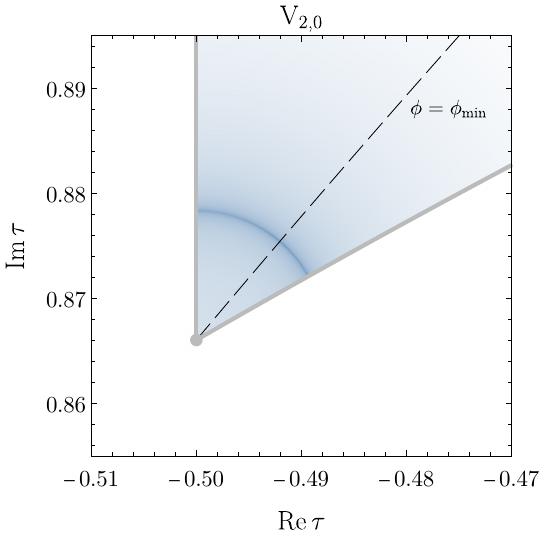} &
    \includegraphics[width=0.42\textwidth]{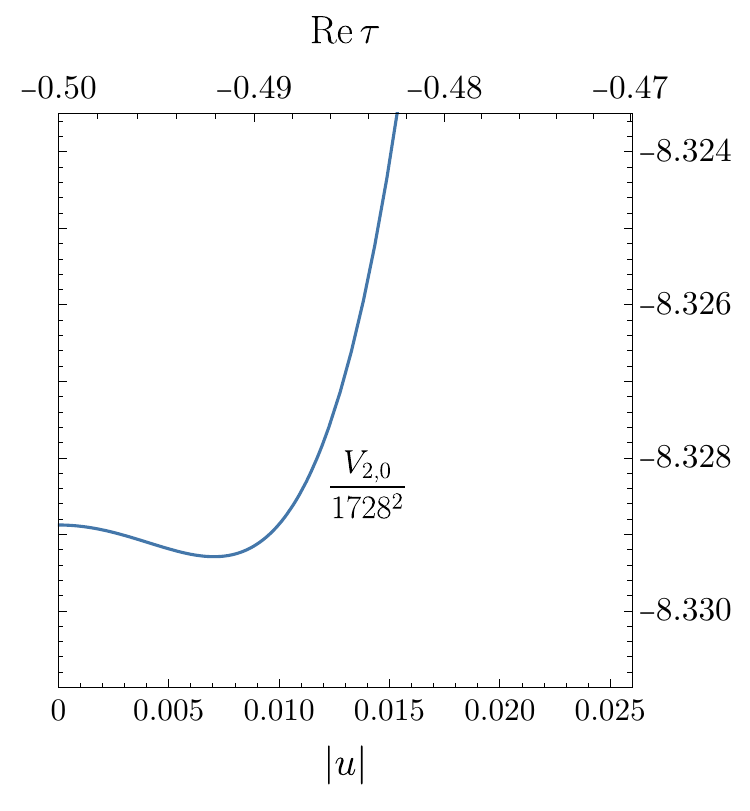}\\[-1ex]
    \includegraphics[width=0.42\textwidth]{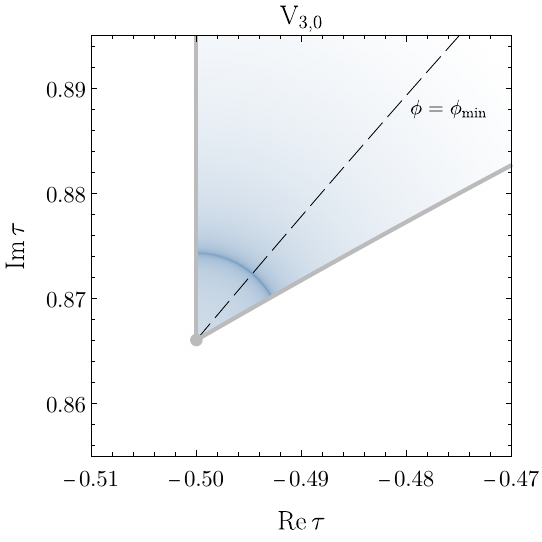} &
    \includegraphics[width=0.42\textwidth]{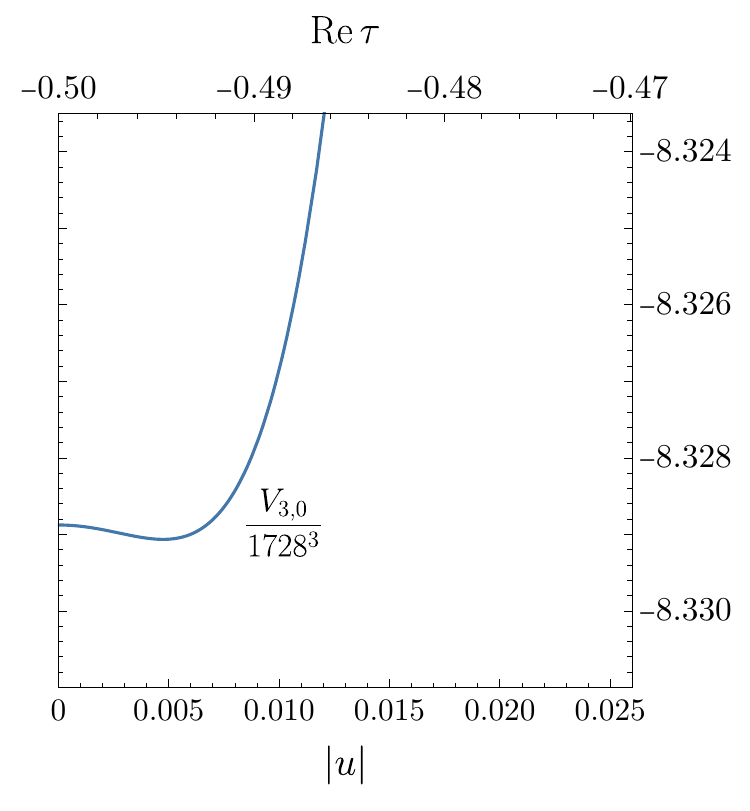}\\[-0.2ex]
    \(\log_{10} \left( \frac{V - V_{\text{min}}}{\lvert V_{\text{min}} \rvert} \right)\)
    \(\vcenter{\hbox{\includegraphics[width=0.28\textwidth]{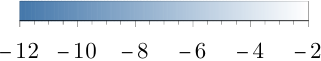}}}\)
  \end{tabular}
  \caption{Potentials \(V_{m,0}(\tau, \bar{\tau})\), \(m = 1, 2, 3\), in the vicinity of the cusp (left panel) and their 1-dimensional projections onto the curve \(\phi = \phi_{\text{min}}\) (right panel), in units of \(\Lambda_V^4\) (see text for details).}
  \label{fig:density}
\end{figure}

Apart from the overall scale, the potential 
\(V_{m,0}\) in~\cref{eq:Vm0uexp4} depends  
on only one parameter --- \(m\), which takes positive integer values.
One can see from~\cref{eq:coefA} that the quartic term coefficient \((A_m^2 - 3)\) is positive for any \(m \geq 1\), so up to \(\mathcal{O}(\lvert u \rvert^6)\) the potential has the well-known Mexican-hat profile, similar to the Higgs potential in the Standard Model (see~\cref{fig:density}).
This clearly indicates that the cusp \(\tau = \omega \leftrightarrow \lvert u \rvert = 0\) is not the minimum.
Instead, this point is a local maximum, while the true minimum is attained at
\begin{equation}
  \label{eq:umin}
 \lvert u \rvert_{\text{min}}
  \simeq (A_m^2 - 3)^{-1/2} \simeq A_m^{-1}
  = \frac{0.0145}{m + 0.0625}\,.
\end{equation}
%
Comparing this approximation with the minima obtained numerically
from the \(q\)-expan\-sions for \(m \leq 7\), we find excellent agreement, 
as shown in~\cref{fig:umin}.

\Cref{eq:umin} has an important phenomenological implication.
In the vicinity of the left cusp, fermion mass matrix entries in modular-invariant theories are proportional to powers of the small parameter \(\epsilon \sim \lvert u \rvert\)~\cite{Novichkov:2021evw}.
With a suitable choice of fermion field representations under the modular group, this leads to a hierarchical mass pattern of the form \((1, \epsilon, \epsilon^2)\) for three generations of fermions.
Hence~\cref{eq:umin} describes possible values of the small parameter responsible for the hierarchy of fermion masses.
In particular, we see that \(\epsilon \sim 0.01\) for small \(m\) which is consistent with the observed mass hierarchy of charged leptons and quarks.

  A model in which hierarchical charged lepton masses and 
  the observed lepton mixing pattern of two large and one small angles 
  are generated naturally without fine-tuning
  in the vicinity of the left cusp was 
  constructed in~\crefext{section}{4.2} of Ref.~\cite{Novichkov:2021evw}. 
  Statistical analysis showed that this \(S^\prime_4\) 
  model is  phenomenologically viable 
  at \(3\sigma\) confidence level for \(\epsilon \in [0.0163, 0.0214]\), \(\epsilon \simeq 2.8 \lvert u \rvert\), 
  independently of the phase of \(u\), 
  with the best fit value of 
  \(\lvert u \rvert \simeq  0.00664\).  
On the other hand,~\cref{eq:umin} yields a series \(\epsilon \simeq 0.0383, 0.0197, 0.0133, \dotsc\) for \(m = 1, 2, 3, \dotsc .\)
  Quite remarkably, choosing \(m = 2 \leftrightarrow \epsilon \simeq 0.0197\) one gets a value of \(\epsilon\) within the phenomenologically allowed range of the model.%
  \footnote{A more careful analysis of the chi-squared function shows that this value of \(\epsilon\) lies in the \(1.1\sigma\) range.}
  In the original construction the small values of \(|u|\) 
  and correspondingly of \(\epsilon\), for which the model is viable, 
  are unexplained.
  Here we find a natural explanation for these small values, which is general and does not rely on the discussed specific model.
   In other words, the potential \(V_{m,0}\) completes the non-fine-tuned model presented in Ref.~\cite{Novichkov:2021evw} by providing a 
   model-independent universal 
   dynamical origin of the smallness of the deviation of \(\tau\) from its symmetric value.%
  \footnote{Although this model was considered in the context of global SUSY, it can be trivially modified to fit into the supergravity framework by shifting modular weights of the fields so that the superpotential carries weight \(-3\) rather than \(0\).}

So far we have not discussed the phase of \(u\) at the minimum, \(\phi_\text{min}\).
It may seem from~\cref{eq:Vm0uexp4} that the potential is independent of \(\phi\), thus having a flat direction.
However, expanding \(V_{m,0}\) to higher orders in \(\lvert u \rvert\) reveals a mild dependence on \(\phi\): up to an overall factor, we have
\begin{equation}
  \label{eq:Vm0uexp8}
  \begin{split}
    V_{m,0} \propto{}& {-1} - 2 \, \lvert u \rvert^2
    + \left( A_m^2 - 3 \right) \lvert u \rvert^4
    + \left( {-4} + 2A_m^2 + B_m^2 \cos 6\phi \right) \lvert u \rvert^6\\
    &+ 2 A_m B_m^2 \cos 3\phi \, \lvert u \rvert^7
    + \left( {-5} + 3A_m^2 + 2B_m^2 \cos 6\phi \right) \lvert u \rvert^8
    + \mathcal{O}(\lvert u \rvert^9) \,,
  \end{split}
\end{equation}
where
\begin{equation}
  \label{eq:coefB}
  \begin{split}
    B_m^2 &\equiv
      \frac{864 \, \lvert \tilde{\eta}_3 \rvert^3}{\pi^6 \, \tilde{\eta}_0^{27}} \, m
      \left[
        \frac{864 \, \lvert \tilde{\eta}_3 \rvert^3}{\pi^6 \, \tilde{\eta}_0^{27}} \, (m-2)
        + \frac{3 \left( 31 \, \tilde{\eta}_3^2 - 10 \tilde{\eta}_0 \tilde{\eta}_6 \right)}{\tilde{\eta}_0 \lvert \tilde{\eta}_3 \rvert}
      \right]
      + \frac{6 \left( 7 \tilde{\eta}_3^2 - 2 \tilde{\eta}_0 \tilde{\eta}_6 \right)}{\tilde{\eta}_0^2}
    \\
    &\simeq 4730.60 \, m^2 - 2069.73 \, m + 33.26 \,.
  \end{split}
\end{equation}
%
Comparing the last expression with \(A_m^2 \simeq 4730.60 \, m^2 + 591.32 \, m + 18.48\), we notice that
\begin{equation}
  B_m \sim A_m \simeq \lvert u \rvert_{\text{min}}^{-1} \,.
\end{equation}
%
This means that in the vicinity of the minimum:
\begin{itemize}
\item terms of order 6 and higher in \(\lvert u \rvert\) are indeed negligible compared to the quadratic and quartic term, which further justifies the validity of approximation~\eqref{eq:Vm0uexp4} for the estimation of \(\lvert u \rvert_{\text{min}}\);
\item the \(\phi\)-dependent parts of \(\mathcal{O}(\lvert u \rvert^6)\) and \(\mathcal{O}(\lvert u \rvert^7)\) terms are comparable, so they are equally important for the estimation of \(\phi_{\text{min}}\);
\item the \(\phi\)-dependent part of \(\mathcal{O}(\lvert u \rvert^8)\) term is negligible compared to the corresponding parts of the two previous terms.
\end{itemize}
We expect that the last condition holds also for higher-order terms, so that the \(\phi\)-dependent contribution to the potential is dominated by
\begin{equation}
\label{eq:phicontr}
B_m^2 \cos 6\phi \, \lvert u \rvert^6 + 2A_m B_m^2 \cos 3\phi \, \lvert u \rvert^7
\propto \cos 6\phi + 2A_m \lvert u \rvert \cos 3\phi
\simeq \cos 6\phi + 2 \cos 3\phi
\end{equation}
%
at \(\lvert u \rvert = \lvert u \rvert_{\text{min}}\).
Expression \eqref{eq:phicontr} is minimised in the region of interest \([-\pi/3, 0]\) at the following unique value of \(\phi\):
\begin{equation}
  \label{eq:phimin}
  \phi_{\text{min}} \simeq -\frac{2\pi}{9}\,,
\end{equation}
%
independently of \(m\), in excellent agreement with the minima 
obtained numerically.

In the case \(m = 2\) relevant for the non-fine-tuned model 
of Ref.~\cite{Novichkov:2021evw}, one gets
\begin{equation}
  u_{\text{min}} \simeq \frac{0.0145}{2+0.0625} \,e^{-2\pi i/9} \,\leftrightarrow\, \tau_{\text{min}} \simeq -0.492 + 0.875 i
\end{equation}
%
(cf.~Table \ref{tab:minima}), which is again consistent with the allowed 
range of \(\tau\) reported in Ref.~\cite{Novichkov:2021evw}.

While~\cref{eq:umin} shows that the minimum deviates from the \emph{symmetric point}, which may be responsible for mass hierarchies,~\cref{eq:phimin} indicates that the minimum deviates also from the \emph{boundary} of the fundamental domain,
providing an origin of CP breaking.
Indeed, \(\phi = 0\) corresponds to the left vertical boundary, while \(\phi = -\pi/3\) corresponds to the arc (see~\cref{app:uexp}), so that the minimum lies in between.
This can be seen in the left panel of~\cref{fig:density}, which shows the potentials \(V_{m,0}\), \(m = 1,2,3\), in the vicinity of the cusp.
To make the potential shapes clearly visible, we use the logarithmic scale \(\log_{10} \big( \frac{V - V_{\text{min}}}{\lvert V_{\text{min}} \rvert} \big)\), where \(V_{\text{min}}\) is the minimum value of the corresponding potential.
As anticipated, the potentials have a deep narrow ``trench'' around \(\lvert u \rvert = \lvert u \rvert_{\text{min}}\), while the dependence on the phase \(\phi\) is almost unnoticeable.
To illustrate further the Mexican-hat shape of the potentials, in the right panel we report their 1-dimensional profiles as one varies \(\lvert u \rvert\) while keeping \(\phi = \phi_{\text{min}}\) fixed (black dashed line in the left panel).

We have seen in this \namecref{sec:results} that the \(V_{m,0}\) potentials have special properties, which are important from the phenomenological viewpoint.
It may seem however that the choice
\begin{equation}
  \label{eq:Hm0}
  H(\tau) = \left( j(\tau) - 1728 \right)^{m/2} \,,
\end{equation}
%
which leads to such potentials, is not distinguished within the more general class~\eqref{eq:H} from the outset, and thus could be considered as a form of tuning: indeed, one could expect to have generically a non-trivial polynomial \(\mathcal{P}(j)\) as well as \(n \neq 0\), which could change the behaviour of \(V\) dramatically.
We would like to point out that the series~\eqref{eq:Hm0} \emph{does} actually play a special role in the full set of \(H(\tau)\) given by~\cref{eq:H}: in fact,~\cref{eq:Hm0} describes a subset of all possible \(H(\tau)\) which vanish only at the symmetric point \(\tau = i\) (which is itself distinguished by modular symmetry).%
\footnote{To prove this, note that \(\mathcal{P}(j(\tau))\) factorises into monomials of the form \((j(\tau) - z)\), \(z \in \mathbb{C}\), and without loss of generality \(z \neq 1728\) (otherwise the monomial can be absorbed into \((j(\tau) - 1728)^{m/2}\) by redefining \(m\)).
  These monomials vanish for some \(\tau \neq i\) in the fundamental domain~\(\mathcal{D}\), since \(j : \mathcal{D} \to \mathbb{C}\) is a bijection and \(j(i) = 1728\) (see~\cref{app:mod_funcs}).
  Therefore, for \(H(\tau)\) to vanish only at \(\tau = i\), \(\mathcal{P}(j(t))\) has to be trivial.
  Finally, since \(j(\omega) = 0\), \(n\) has to be zero, as otherwise \(H(\tau)\) vanishes at \(\tau = \omega\).}
In view of this special property, we expect that \(V_{m,0}\) potentials should arise naturally in certain top-down completions without any need for special tuning to avoid non-trivial \(\mathcal{P}\) and non-zero \(n\).

\section{Summary and Conclusions}
\label{sec:summary}

In the present article we have investigated the problem of modulus 
stabilisation in theories of flavour based on modular symmetry.
The modulus \(\tau\) --- a complex scalar field --- 
plays a fundamental role in the modular-invariance approach 
to the lepton and quark flavour problems. 
It has specific transformation properties under the action of the 
modular group \(\Gamma \equiv SL(2,\mathbb{Z})\). 
The VEV of the modulus \(\tau\) 
can be the only source of breaking of both the modular symmetry 
and the flavour symmetry, described in the 
approach by a finite inhomogeneous (homogeneous) 
modular group \(\Gamma^{(\prime)}_N\). Thus, 
flavons are not needed. 
For \(N\leq 5\), the finite modular groups \(\Gamma_N\) 
are isomorphic to the permutation groups 
\( S_3\), \( A_4\), \( S_4\) and \( A_5\), while 
\(\Gamma^\prime_N\) are isomorphic to their double covers. 
In the ``minimal'' models without flavons,  
the VEV of \(\tau\) can also be the only source 
of breaking of CP symmetry when it does not lie 
on the imaginary axis \(\re \tau = 0\) or on the border
of the fundamental domain \(\mathcal{D}\) of the modular group,
where it has CP-conserving values. 
In the discussed approach to the flavour problem, 
the elements of the Yukawa coupling 
and fermion mass matrices in the Lagrangian
are expressed in terms of modular forms of a certain level \(N\) 
and a limited number of coupling constants.
The modular forms are functions of the modulus \(\tau\),
have specific transformation properties under the action of 
the modular group \(SL(2,\mathbb{Z})\)
and furnish irreducible representations of the 
finite modular, i.e., flavour symmetry, group \(\Gamma^{(\prime)}_N\).
The matter fields are assumed also to transform 
in representations of \(\Gamma^{(\prime)}_N\).
After the flavour symmetry is (fully or partially) broken 
by the VEV of \(\tau\), the modular forms and thus 
the elements of the Yukawa coupling and fermion mass matrices get fixed. 
Correspondingly, the fermion mass matrices exhibit a certain 
symmetry-determined flavour structure which depends 
via the modular forms used on the VEV.
Thus, \(\tau\)'s VEV is critical for 
the phenomenological viability of a given modular-invariant 
flavour model. 

Although there is no VEV of \(\tau\) 
which preserves the full modular symmetry,
there exist three values  
in the modular group fundamental domain, which 
break the modular symmetry only partially  
\cite{Novichkov:2018ovf}.
Only one of these residual symmetry points is relevant 
for our analysis, namely, 
\(\tau_\text{sym} = \omega  \equiv \exp(2\pi i/ 3) 
 = -1/2 + \sqrt{3}/2 \,i\) (the ``left cusp''), 
at which the \(\mathbb{Z}^{ST}_3\) symmetry is preserved. 
In models where \(\tau\) deviates slightly from 
\(\tau_\text{sym} = \omega\), charged lepton (and possibly quark) 
mass hierarchies may arise naturally 
from the properties of the modular forms 
as powers of the small deviation 
\(\lvert \tau-\tau_\text{sym} \rvert\)  
without the use of fine-tuned constants~\cite{Novichkov:2021evw}.

 Following a bottom-up approach, a large number of 
viable ``minimal''  lepton and quark flavour models 
based on modular symmetry, which do not include flavons, 
has been constructed. In the overwhelming majority of these models 
the VEV of the modulus has been determined by confronting 
model predictions with experimental 
data and can vary significantly depending on the model.
There have been a few attempts to determine the modulus VEV 
from a dynamical principle (see~\cref{sec:intro}). 
The results of these attempts revealed, in particular, that in the 
predominant number of cases the specific CP-invariant potentials 
used for \(\tau\) lead  to CP-conserving VEVs of \(\tau\). 

In the present study of  modulus stabilisation 
we have considered relatively simple UV-motivated 
modular- and CP-invariant potentials of the modulus, 
\(V_{m,n}(\tau,\overline{\tau})\) (\cref{eq:V}), 
\(m,n\) being non-negative integer numbers,
proposed and analysed within the framework of 
supergravity theories in Ref.~\cite{Cvetic:1991qm} and 
further studied in Ref.~\cite{Gonzalo:2018guu}.
They can be expressed in terms of the Dedekind eta 
function, \(\eta(\tau)\), and its  derivatives. 
Using the well-known \(q\)-expansion of \(\eta(\tau)\) 
(\cref{eq:etaq0}),  
which allows to compute the potential 
\(V_{m,n}(\tau,\overline{\tau})\) to arbitrary precision in any point 
of the fundamental domain, we have 
derived the absolute minima of \(V_{m,n}\) for 
\(m,n = 0,1,2,3\) 
(\cref{tab:minima,fig:summary})
and any \(m > 0\) for \(n=0\)
(\cref{sec:CPVminima}).
It was conjectured in Ref.~\cite{Cvetic:1991qm}
that all extrema of \(V_{m,n}\) would correspond to CP-conserving values of \(\tau\), i.e., would lie either on the boundary of the 
fundamental domain \(\mathcal{D}\) or on the imaginary axis. 
In~\cite{Cvetic:1991qm} the cases \((m,n) = (0,0), (1,1), (0,3)\) 
were explicitly examined and the global minima of the corresponding 
potentials were indeed found 
to lie at \(\tau \simeq 1.2\, i\) (imaginary axis), \(\tau \simeq \pm 0.24+0.97\,i\) (equivalent minima on the unit arc) and \(\tau = i\), respectively.
While we have verified these results we showed also that 
i) the potentials \(V_{0,n}\) for \(n=1,2\) have the same absolute 
minimum as \(V_{0,3}\),
ii) the potentials \(V_{m,n}\) with \(m,n =1,2,3\),  
have absolute minima at the unit arc, which shift towards 
\(\tau = \omega\) (\(\tau = i\)) along the arc as \(m\) (\(n\)) grows.  
Most importantly, we have further found that
potentials with \(n=0\) but given \(m>0\) do allow for 
a pair of degenerate (CP-conjugate) global minima 
at \(\tau_\text{min}\) and \((-\overline{\tau}_\text{min})\), 
which break CP symmetry spontaneously. 
Moreover, 
\(\tau_\text{min}\) are found to be located in the vicinity of the 
left cusp \(\tau = \omega\) 
(\cref{fig:density}), at values of \(\lvert \tau_\text{min}-\omega \rvert\) 
favoured by the mechanism put forward in~\cite{Novichkov:2021evw} to explain fermion (charged-lepton and quark) mass hierarchies.
As the found CP-breaking minima deviate only slightly from the 
fundamental domain boundary,  to make sure these deviations are not 
a numerical artefact of the used \(q\)-expansions of \(\eta(\tau)\),
we re-expanded \(\eta(\tau)\) and the potential \(V_{m,0}\) in 
terms of the parameter \(u = (\tau - \omega)/(\tau - \omega^2)\) 
which quantifies the deviation of \(\tau\) from the left cusp
(\cref{sec:quexpansions}). 
Expressed in terms of \(u\) this potential is shown
to depend, apart from the overall scale, 
on just one parameter --- \(m\), which takes 
positive integer values.
We found that up to \(\mathcal{O}(\lvert u \rvert^6)\) 
(with \(\lvert u \rvert^6\) giving negligible contribution),
the potential \(V_{m,0}\) has the well-known Mexican-hat profile, 
similar to the Higgs potential in the Standard Model 
(\cref{eq:Vm0uexp4}), with absolute minimum 
at \(\lvert u \rvert_{\text{min}}  
\simeq 0.0145/(m+0.0625)\) --- 
in excellent agreement with the minima obtained numerically from 
the \(q\)-expansions for \(m \leq 7\) (\cref{fig:umin}).
Using further the expansions up to \(\mathcal{O}(\lvert u \rvert^8)\) 
(with \(\lvert u \rvert^8\) being negligibly small) 
we have found also 
that \(\arg(u_{\text{min}}) \simeq -2\pi/9 = -40^\circ\) independently of \(m\).

 An \(S^\prime_4\) lepton flavour model   
constructed in~\cite{Novichkov:2021evw}, in which
the charged lepton mass hierarchies are generated naturally for \(\tau\) 
in the vicinity of the left cusp and the observed lepton mixing 
is reproduced without fine-tuning, 
was found to be phenomenologically viable at \(3\sigma\) 
C.L.~for \(\epsilon \simeq 2.8 \lvert u \rvert \in [0.0163, 0.0214]\).
It is quite remarkable that for \(m=2\), the potential \(V_{2,0}\) has 
an absolute minimum 
at \(\lvert u \rvert_{\text{min}} \simeq 0.00705\) 
corresponding to \(\epsilon \simeq 0.0197\)
lying in the \({\sim}1\sigma\) allowed range of the model. Thus,
the potential \(V_{2,0}\) completes this non-fine-tuned 
lepton flavour model 
by providing a dynamical origin of the smallness of the deviation of 
\(\tau\) from its left cusp symmetric value.

 We note finally that the results of our study 
of modulus stabilisation 
do not depend on the choice of the finite modular 
group \(\Gamma^{(\prime)}_N\) 
as a group of flavour symmetry, of the modular weights 
of the matter fields and of the representations  of 
\(\Gamma^{(\prime)}_N\) assumed to be furnished by the matter fields,
which define a modular-invariant model of flavour.
In this sense they are universal.
They have a direct impact on the phenomenology 
of the modular-invariant models of flavour since they 
lay out preferred regions in the fundamental domain 
of the modular group for stabilisation of the modulus. 
Our results may have also implications for the problem of CP violation in supersymmetric extensions of the 
Standard Model.

\vspace{0.5cm}
\section*{Acknowledgements}

This project has received funding/support from the European Union's Horizon 2020 research and innovation programme under the Marie Skłodowska-Curie grant agreement No.~860881-HIDDeN.
This work was supported in part
by the INFN program on Theoretical Astroparticle Physics (P.P.N. and S.T.P.)
and by the  World Premier International Research Center
Initiative (WPI Initiative, MEXT), Japan (S.T.P.).
P.P.N.'s work was supported in part by the European Research Council, under grant ERC-AdG-885414.
P.P.N. would like to thank the Astroparticle Physics sector of SISSA for hospitality and support.
The work of J.T.P.~was supported by
Fundação para a Ciência e a Tecnologia (FCT, Portugal) through the projects
PTDC/FIS-PAR/29436/2017, 
CERN/FIS-PAR/0004/2019, CERN/FIS-PAR/0008/2019, and
CFTP-FCT Unit 777 (namely UIDB/00777/2020 and UIDP/00777/2020),
which are partially funded through POCTI (FEDER), COMPETE, QREN and EU.
  
\appendix
\section{Modular forms}
\label{app:mod_funcs}

The \emph{Dedekind eta function} is a modular form of weight \(1/2\) defined as
\begin{equation}
    \eta(\tau) \equiv q^{1/24} \prod_{n=1}^{\infty}\left(1 - q^n\right) =  q^{1/24}\, \phi(q)\,,
\end{equation}
where \(q \equiv e^{2\pi i\tau}\) and \(\phi(q)\) is known as the Euler function.
The eta function admits the expansions
\begin{equation}
\begin{aligned}
        \eta 
        &= q^{1/24} \left(1 - q - q^{2} + q^{5} + q^{7} - q^{12} - q^{15} + \mathcal{O}(q^{22})\right)
        \\
        &= q^{1/24} \left(1+q+2q^2+3q^3+5q^4+7q^5+11q^6+\mathcal{O}(q^{7})\right)^{-1}
        \,,
\end{aligned}
\label{eq:etaq}
\end{equation}
and satisfies \(\eta(T\tau)=\eta(\tau+1)= e^{i \pi/12}\,\eta(\tau)\) and \(\eta(S\tau)=\eta(-1/\tau)=\sqrt{-i\tau}\,\eta(\tau)\).

\vskip 4mm

The \emph{Eisenstein series of weight \(2k\)} is defined for integer \(k>1\) as
\begin{align}
G_{2k}(\tau) &= \sum_{\substack{n_1,n_2\in \mathbb{Z}\\(n_1,n_2)\neq (0,0)}}(n_1+n_2 \tau)^{-2k}\,,
\label{eq:eisenstein}
\end{align}
and converges to a holomorphic function in the upper-half plane: a modular form of weight \(2k\). 
While the series does not converge for \(k=1\), one can still define the \(G_2(\tau)\) function via a specific prescription on the order of summation
(see, e.g.,~\cite{Schoeneberg:1974md}, where it is denoted \(G_2^*\)).
This function is related to \(G_4\) by the identity~\cite{Cvetic:1991qm}
\begin{equation}
\frac{5}{2\pi} G_4 = i\, G_2' + \frac{G_2^2}{2\pi}\,.
\label{eq:G4G2}
\end{equation}
Using~\cref{eq:G2}, one can further show that \(G_2\) is not quite a modular form of weight 2, since under a generic modular transformation \(\gamma \in SL(2,\mathbb{Z})\) 
\begin{equation}
   \frac{\eta'(\tau)}{\eta(\tau)} \,\xrightarrow{\gamma}\, 
   (c\tau + d)^2 \, \frac{\eta'(\tau)}{\eta(\tau)} + \frac{1}{2}\,c(c\tau +d)\,.
\end{equation}
Noting how \(\im\tau\) transforms under the action of the modular group, it further follows that
\begin{equation}
\frac{1}{4i\,\im\tau} \,\xrightarrow{\gamma}\,
\frac{|c\tau + d|^2}{4i\,\im\tau} = 
\frac{(c\tau + d)^2}{4i\,\im\tau} 
- \frac{1}{2}\, c(c\tau + d)\,.
\end{equation}
One can then define \(\hat{G}_2\) as given in~\cref{eq:G2hat}, which transforms as a weight 2 form, at the cost of being non-holomorphic.

\vskip 4mm
Finally, the \emph{Klein \(j\) function} or \emph{\(j\)-invariant} (sometimes called the absolute modular invariant) is a modular form of zero weight.
It can be defined in terms of the Dedekind eta
and \(G_4\) as
\begin{equation}
j(\tau) \equiv \frac{3^6 5^3}{\pi^{12}} \frac{G_4(\tau)^3}{\eta(\tau)^{24}}\,.
\end{equation}
Using~\cref{eq:G2,eq:G4G2}, one can relate the \(j\) function to the Dedekind eta and its derivatives, showing that
\begin{equation}
  j = \left( \frac{72}{\pi^2} \, \frac{\eta \eta'' - 3 \eta'^2}{\eta^{10}} \right)^3
  = \left[
  \frac{72}{\pi^2 \eta^6} \left( \frac{\eta'}{\eta^3} \right)' \,
  \right]^3 \,,
\end{equation}
as given in~\cref{eq:jmain}.
This function  \(j : \mathcal{D} \to \mathbb{C}\) is a one-to-one map between points in the fundamental domain and the whole complex plane. In particular, \(j(\tau)\) takes real values only for CP-conserving values of \(\tau\), i.e.,~when \(\tau\) is on the border of \(\mathcal{D}\) or when \(\re\tau=0\).
The \(j\) function is not holomorphic at \(\tau= i\infty\), where it diverges. 
At the remaining symmetric points, one has \(j(\omega) = 0\) and \(j(i) = 1728= 12^3\).
Thus, it admits as \(q\)-expansion the Laurent series
\begin{equation}
j =
744 + \frac{1}{q} + 196884\, q + 21493760\, q^2 + 864299970\, q^3  
 + \mathcal{O}(q^4)
\,.
\end{equation}
While here only the first terms in this well-known expansion are reported, in practice many more powers of \(q\) were taken into account in our analyses, guaranteeing numerical convergence and stability.
It is further known that \(j\) has a triple zero at \(\tau=\omega\), that \((j-1728)\) has a double zero at \(\tau = i\) and that the derivative \(j'\) vanishes only at these values of \(\tau\): \(j'(\omega)= j'(i) = 0\) (see, e.g.,~\cite{Apostol:1990mf}).

\section{Rigid SUSY limit}
\label{app:rigid}

In the limit of rigid \(\mathcal{N}=1\) SUSY, one has \(M_P \to \infty \) and \(\kappa \to 0\).
Keeping the form of \(K(\tau,\overline\tau)\) given in~\cref{eq:Kahler}, 
one sees that the scalar potential of~\cref{eq:V1} becomes
\begin{equation}
V = K^{i \bar{j}} \partial_i W \partial_{\bar{j}} W^* \,.
\end{equation}
Note that, in this limit,
\(\mathfrak{n} = \kappa^2 \Lambda_K^2 \to 0\).
The superpotential of~\cref{eq:superpotential} reduces to \(W(\tau)  = \Lambda_W^3 H(\tau)\) and we arrive at the simple result
\begin{equation}
    V(\tau,\overline{\tau}) = \frac{4\Lambda_W^6}{\Lambda_K^2} (\im\tau)^2 \left|H'(\tau)\right|^2\,.
\end{equation}
Taking \(H\) of the form given in~\cref{eq:H}, with \(\mathcal{P}(j)=1\),
one finds that there is always a value of \(\tau \in \mathcal{D}\)
for which \(H'(\tau)= 0\). Hence,
global minima of this potential
correspond to the zeros of \(H'\).
This function is given by
\begin{equation}
    H' = j' \left(j-1728\right)^{m/2} j^{n/3} \left[\frac{m}{2} \frac{1}{j-1728}+\frac{n}{3} \frac{1}{j}\right]\,.
\end{equation}
Apart from the trivial case \(m=n=0\),
the zeros of \(H'\) --- and correspondingly the global minima of \(V\) --- are located at CP-conserving values of \(\tau\), namely at \(\tau=i\), \(\tau=\omega\), or at points for which the factor in square brackets vanishes, which correspond to 
real \(j\in[0,1728]\) and values of \(\tau\) on the arc.
This result suggests that supergravity effects are important for the presence of CP-violating minima in the discussed class of simple superpotentials.
More specifically, the presence of the term \(\propto \hat{G}_2\) and the term 
\(\propto 3|H|^2\) in the potential \(V\) in \cref{eq:V2,eq:V}
seem to be crucial for the spontaneous breaking of the CP symmetry.
Let us further note that,
in the case of a non-trivial \(\mathcal{P}(j)\), this polynomial can be engineered to produce minima at arbitrary points in the fundamental domain.

\section{\texorpdfstring{\(u\)-expansions}{u-expansions}}
\label{app:uexp}

Recalling the definition of \(u\) given in~\cref{sec:quexpansions},
\begin{equation}
u\equiv \frac{\tau-\omega}{\tau-\omega^2} 
\quad \Leftrightarrow \quad \tau = \omega^2\, \frac{\omega^2 -u}{1-u}\,,  
\end{equation}
one finds
\begin{equation}
\re\tau \,=\, -\, \dfrac{1}{2} -\, \dfrac{\sqrt{3}\,\im u}{|1 - u|^2}\,,
\qquad
\im\tau \,=\, \dfrac{\sqrt{3}}{2}\,\dfrac{1 - |u|^2}{|1 - u|^2}\,,
\end{equation}
and conversely
\begin{equation}
\re u \,=\, 
\frac{\re\tau+|\tau|^2-1/2}{|\tau-\omega^2|^2}\,,
\qquad
\im u \,=\, -\frac{\sqrt{3}}{2}\,\frac{1+2 \re\tau}{|\tau-\omega^2|^2}\,.
\end{equation}
%
Writing \(u = |u|e^{i\phi}\), as in~\cref{sec:CPVminima},
one can check that \(\re u > 0\) and
\begin{equation}
    \phi = - \arctan\left(\frac{\sqrt{3}}{2}\,\frac{1+2 \re \tau}{\re\tau+|\tau|^2-1/2}\right)
\end{equation}
within the fundamental domain (excluding \(\tau = \omega\), where \(u=0\) and \(\phi\) is indeterminate).
By analysing this expression, it follows that the phase of \(u\) varies in the interval \([-\pi/3,0]\). Namely, it reaches its highest value of \(\phi=0\) at the left boundary of the fundamental domain, \(\re\tau=-1/2\). Its lowest value corresponds to the maximum value of the argument of the arctangent, attained at the arc \(|\tau|^2 = 1\), for which \(\phi = -\arctan\sqrt{3} = - \pi/3\).

\vskip 4mm

Following the procedure described in~\cref{sec:quexpansions}, one can obtain the \(u\)-expansions of modular forms, such as that of \(\eta = \sqrt{1-u}\,\tilde\eta\), which can be extracted from~\cref{eq:tildeuexp}.
For \(j\), \(\tilde{G_2}\) and \(\tilde{G_4}\),
having defined \(\tilde G_2\) and \(\tilde G_4\) via
\begin{equation}
\begin{aligned}
G_2 (u) &\equiv \frac{2\pi}{\sqrt{3}}\left((1-u)+(1-u)^2\,\tilde{G}_2\right)\,,
\\[2mm]
{G}_4(u)&\equiv (1-u)^{4}\, \tilde{G}_4(u) \,,
\end{aligned}
\end{equation}
we find
\begin{equation}
\begin{aligned}
  j(u) &\,\simeq\, -237698\, u^3 - 1.17505 \times 10^7\, u^6 - 2.78879 \times 10^8\, u^9 + \mathcal{O}(u^{12}) \,,
\\[2mm]  
\tilde{G}_2(u) 
&\,\simeq\, 4.29865\, u^2 +14.7827\, u^5 + 18.155977\, u^8 + \mathcal{O}(u^{11}) \,,
\\[2mm]  
\tilde{G}_4(u) 
&\,\simeq\, 22.6272\, u + 243.166\, u^4 + 716.769\, u^7+ \mathcal{O}(u^{10})\,.
\end{aligned}
\label{eq:uexps}
\end{equation}
Note that, while both \(j\) and \(\tilde\eta\) are invariant under \(\gamma = ST\), one can check that \(\tilde G_2 \xrightarrow{ST} \omega\, \tilde G_2\) and \(\tilde G_4 \xrightarrow{ST} \omega^2\, \tilde G_4\). Recalling that \(u\xrightarrow{ST} \omega^2\, u\) and \(j(\omega)=0\), the peculiar structure in powers of \(u\) of~\cref{eq:uexps} follows.

\bibliographystyle{JHEPwithnote}
\bibliography{bibliography}

\providecommand{\noopsort}[1]{}\providecommand{\singleletter}[1]{#1}%

\providecommand{\href}[2]{#2}\begingroup\raggedright\begin{thebibliography}{10}

\bibitem{Feruglio:2015jfa}
F.~Feruglio, \emph{{Pieces of the Flavour Puzzle}},
  \href{https://doi.org/10.1140/epjc/s10052-015-3576-5}{\emph{Eur. Phys. J. C}
  {\bfseries 75} (2015) 373}
  [\href{https://arxiv.org/abs/1503.04071}{{\ttfamily 1503.04071}}].

\bibitem{PDG2019}
K.~Nakamura and S.~T. Petcov, \emph{{Neutrino Masses, Mixing, and
  Oscillations\emph{, in M. Tanabashi et al. (Particle Data Group),} Review of
  Particle Physics}},
  \href{https://doi.org/10.1103/PhysRevD.98.030001}{\emph{Phys. Rev.}
  {\bfseries D98} (2018) 030001} and 2019 update.

\bibitem{Fukuda:1998mi}
{\scshape Super-Kamiokande} collaboration, Y.~Fukuda et~al., \emph{{Evidence
  for oscillation of atmospheric neutrinos}},
  \href{https://doi.org/10.1103/PhysRevLett.81.1562}{\emph{Phys. Rev. Lett.}
  {\bfseries 81} (1998) 1562}
  [\href{https://arxiv.org/abs/hep-ex/9807003}{{\ttfamily hep-ex/9807003}}].

\bibitem{Feruglio:2017spp}
F.~Feruglio, \emph{{Are neutrino masses modular forms?}},  in \emph{From My
  Vast Repertoire...: Guido Altarelli's Legacy} (A.~Levy, S.~Forte and
  G.~Ridolfi, eds.), pp.~227--266.
\newblock World Scientific Publishing, 2019.
\newblock [\href{https://arxiv.org/abs/1706.08749}{{\ttfamily 1706.08749}}].

\bibitem{Kobayashi:2018vbk}
T.~Kobayashi, K.~Tanaka and T.~H. Tatsuishi, \emph{{Neutrino mixing from finite
  modular groups}},
  \href{https://doi.org/10.1103/PhysRevD.98.016004}{\emph{Phys.\ Rev.\ D}
  {\bfseries 98} (2018) 016004}
  [\href{https://arxiv.org/abs/1803.10391}{{\ttfamily 1803.10391}}].

\bibitem{Penedo:2018nmg}
J.~T. Penedo and S.~T. Petcov, \emph{{Lepton Masses and Mixing from Modular
  $S_4$ Symmetry}},
  \href{https://doi.org/10.1016/j.nuclphysb.2018.12.016}{\emph{Nucl. Phys.}
  {\bfseries B939} (2019) 292}
  [\href{https://arxiv.org/abs/1806.11040}{{\ttfamily 1806.11040}}].

\bibitem{Criado:2018thu}
J.~C. Criado and F.~Feruglio, \emph{{Modular Invariance Faces Precision
  Neutrino Data}},
  \href{https://doi.org/10.21468/SciPostPhys.5.5.042}{\emph{SciPost Phys.}
  {\bfseries 5} (2018) 042} [\href{https://arxiv.org/abs/1807.01125}{{\ttfamily
  1807.01125}}].

\bibitem{deAdelhartToorop:2011re}
R.~de~Adelhart~Toorop, F.~Feruglio and C.~Hagedorn, \emph{{Finite Modular
  Groups and Lepton Mixing}},
  \href{https://doi.org/10.1016/j.nuclphysb.2012.01.017}{\emph{Nucl. Phys. B}
  {\bfseries 858} (2012) 437}
  [\href{https://arxiv.org/abs/1112.1340}{{\ttfamily 1112.1340}}].

\bibitem{Altarelli:2010gt}
G.~Altarelli and F.~Feruglio, \emph{{Discrete Flavor Symmetries and Models of
  Neutrino Mixing}},
  \href{https://doi.org/10.1103/RevModPhys.82.2701}{\emph{Rev. Mod. Phys.}
  {\bfseries 82} (2010) 2701}
  [\href{https://arxiv.org/abs/1002.0211}{{\ttfamily 1002.0211}}].

\bibitem{Ishimori:2010au}
H.~Ishimori, T.~Kobayashi, H.~Ohki, Y.~Shimizu, H.~Okada and M.~Tanimoto,
  \emph{{Non-Abelian Discrete Symmetries in Particle Physics}},
  \href{https://doi.org/10.1143/PTPS.183.1}{\emph{Prog. Theor. Phys. Suppl.}
  {\bfseries 183} (2010) 1} [\href{https://arxiv.org/abs/1003.3552}{{\ttfamily
  1003.3552}}].

\bibitem{King:2014nza}
S.~F. King, A.~Merle, S.~Morisi, Y.~Shimizu and M.~Tanimoto, \emph{{Neutrino
  Mass and Mixing: from Theory to Experiment}},
  \href{https://doi.org/10.1088/1367-2630/16/4/045018}{\emph{New J. Phys.}
  {\bfseries 16} (2014) 045018}
  [\href{https://arxiv.org/abs/1402.4271}{{\ttfamily 1402.4271}}].

\bibitem{Tanimoto:2015nfa}
M.~Tanimoto, \emph{{Neutrinos and flavor symmetries}},
  \href{https://doi.org/10.1063/1.4915578}{\emph{AIP Conf. Proc.} {\bfseries
  1666} (2015) 120002}.

\bibitem{Petcov:2017ggy}
S.~T. Petcov, \emph{{Discrete Flavour Symmetries, Neutrino Mixing and Leptonic
  CP Violation}},
  \href{https://doi.org/10.1140/epjc/s10052-018-6158-5}{\emph{Eur. Phys. J.}
  {\bfseries C78} (2018) 709}
  [\href{https://arxiv.org/abs/1711.10806}{{\ttfamily 1711.10806}}].

\bibitem{Novichkov:2019sqv}
P.~P. Novichkov, J.~T. Penedo, S.~T. Petcov and A.~V. Titov, \emph{{Generalised
  CP Symmetry in Modular-Invariant Models of Flavour}},
  \href{https://doi.org/10.1007/JHEP07(2019)165}{\emph{JHEP} {\bfseries 07}
  (2019) 165} [\href{https://arxiv.org/abs/1905.11970}{{\ttfamily
  1905.11970}}].

\bibitem{Novichkov:2018ovf}
P.~Novichkov, J.~Penedo, S.~Petcov and A.~Titov, \emph{{Modular S$_{4}$ models
  of lepton masses and mixing}},
  \href{https://doi.org/10.1007/JHEP04(2019)005}{\emph{JHEP} {\bfseries 04}
  (2019) 005} [\href{https://arxiv.org/abs/1811.04933}{{\ttfamily
  1811.04933}}].

\bibitem{Novichkov:2018yse}
P.~Novichkov, S.~Petcov and M.~Tanimoto, \emph{{Trimaximal Neutrino Mixing from
  Modular A4 Invariance with Residual Symmetries}},
  \href{https://doi.org/10.1016/j.physletb.2019.04.043}{\emph{Phys.\ Lett.\ B}
  {\bfseries 793} (2019) 247}
  [\href{https://arxiv.org/abs/1812.11289}{{\ttfamily 1812.11289}}].

\bibitem{Novichkov:2018nkm}
P.~Novichkov, J.~Penedo, S.~Petcov and A.~Titov, \emph{{Modular A$_{5}$
  symmetry for flavour model building}},
  \href{https://doi.org/10.1007/JHEP04(2019)174}{\emph{JHEP} {\bfseries 04}
  (2019) 174} [\href{https://arxiv.org/abs/1812.02158}{{\ttfamily
  1812.02158}}].

\bibitem{Okada:2020brs}
H.~Okada and M.~Tanimoto, \emph{{Spontaneous CP violation by modulus $\tau$ in
  $A_4$ model of lepton flavors}},
  \href{https://doi.org/10.1007/JHEP03(2021)010}{\emph{JHEP} {\bfseries 03}
  (2021) 010} [\href{https://arxiv.org/abs/2012.01688}{{\ttfamily
  2012.01688}}].

\bibitem{Novichkov:2020eep}
P.~P. Novichkov, J.~T. Penedo and S.~T. Petcov, \emph{{Double cover of modular
  $S_4$ for flavour model building}},
  \href{https://doi.org/10.1016/j.nuclphysb.2020.115301}{\emph{Nucl. Phys. B}
  {\bfseries 963} (2021) 115301}
  [\href{https://arxiv.org/abs/2006.03058}{{\ttfamily 2006.03058}}].

\bibitem{Novichkov:2021evw}
P.~P. Novichkov, J.~T. Penedo and S.~T. Petcov, \emph{{Fermion mass
  hierarchies, large lepton mixing and residual modular symmetries}},
  \href{https://doi.org/10.1007/JHEP04(2021)206}{\emph{JHEP} {\bfseries 04}
  (2021) 206} [\href{https://arxiv.org/abs/2102.07488}{{\ttfamily
  2102.07488}}].

\bibitem{Kobayashi:2019mna}
T.~Kobayashi, Y.~Shimizu, K.~Takagi, M.~Tanimoto and T.~H. Tatsuishi,
  \emph{{New $A_4$ lepton flavor model from $S_4$ modular symmetry}},
  \href{https://doi.org/10.1007/JHEP02(2020)097}{\emph{JHEP} {\bfseries 02}
  (2020) 097} [\href{https://arxiv.org/abs/1907.09141}{{\ttfamily
  1907.09141}}].

\bibitem{Kobayashi:2019xvz}
T.~Kobayashi, Y.~Shimizu, K.~Takagi, M.~Tanimoto and T.~H. Tatsuishi,
  \emph{{$A_4$ lepton flavor model and modulus stabilization from $S_4$ modular
  symmetry}}, \href{https://doi.org/10.1103/PhysRevD.100.115045}{\emph{Phys.
  Rev. D} {\bfseries 100} (2019) 115045}
  [\href{https://arxiv.org/abs/1909.05139}{{\ttfamily 1909.05139}}], [Erratum:
  Phys.Rev.D 101, 039904 (2020)].

\bibitem{Gui-JunDing:2019wap}
G.-J. Ding, S.~F. King, X.-G. Liu and J.-N. Lu, \emph{{Modular S$_{4}$ and
  A$_{4}$ symmetries and their fixed points: new predictive examples of lepton
  mixing}}, \href{https://doi.org/10.1007/JHEP12(2019)030}{\emph{JHEP}
  {\bfseries 12} (2019) 030}
  [\href{https://arxiv.org/abs/1910.03460}{{\ttfamily 1910.03460}}].

\bibitem{Ding:2019xna}
G.-J. Ding, S.~F. King and X.-G. Liu, \emph{{Neutrino mass and mixing with
  $A_5$ modular symmetry}},
  \href{https://doi.org/10.1103/PhysRevD.100.115005}{\emph{Phys.\ Rev.\ D}
  {\bfseries 100} (2019) 115005}
  [\href{https://arxiv.org/abs/1903.12588}{{\ttfamily 1903.12588}}].

\bibitem{Kobayashi:2018scp}
T.~Kobayashi, N.~Omoto, Y.~Shimizu, K.~Takagi, M.~Tanimoto and T.~H. Tatsuishi,
  \emph{{Modular A$_{4}$ invariance and neutrino mixing}},
  \href{https://doi.org/10.1007/JHEP11(2018)196}{\emph{JHEP} {\bfseries 11}
  (2018) 196} [\href{https://arxiv.org/abs/1808.03012}{{\ttfamily
  1808.03012}}].

\bibitem{Ding:2019zxk}
G.-J. Ding, S.~F. King and X.-G. Liu, \emph{{Modular A$_{4}$ symmetry models of
  neutrinos and charged leptons}},
  \href{https://doi.org/10.1007/JHEP09(2019)074}{\emph{JHEP} {\bfseries 09}
  (2019) 074} [\href{https://arxiv.org/abs/1907.11714}{{\ttfamily
  1907.11714}}].

\bibitem{Kobayashi:2019gtp}
T.~Kobayashi, T.~Nomura and T.~Shimomura, \emph{{Type II seesaw models with
  modular $A_4$ symmetry}},
  \href{https://doi.org/10.1103/PhysRevD.102.035019}{\emph{Phys. Rev. D}
  {\bfseries 102} (2020) 035019}
  [\href{https://arxiv.org/abs/1912.00637}{{\ttfamily 1912.00637}}].

\bibitem{Okada:2019xqk}
H.~Okada and Y.~Orikasa, \emph{{Modular $S_3$ symmetric radiative seesaw
  model}}, \href{https://doi.org/10.1103/PhysRevD.100.115037}{\emph{Phys.\
  Rev.\ D} {\bfseries 100} (2019) 115037}
  [\href{https://arxiv.org/abs/1907.04716}{{\ttfamily 1907.04716}}].

\bibitem{Ding:2020msi}
G.-J. Ding, S.~F. King, C.-C. Li and Y.-L. Zhou, \emph{{Modular Invariant
  Models of Leptons at Level 7}},
  \href{https://doi.org/10.1007/JHEP08(2020)164}{\emph{JHEP} {\bfseries 08}
  (2020) 164} [\href{https://arxiv.org/abs/2004.12662}{{\ttfamily
  2004.12662}}].

\bibitem{Okada:2018yrn}
H.~Okada and M.~Tanimoto, \emph{{CP violation of quarks in $A_4$ modular
  invariance}},
  \href{https://doi.org/10.1016/j.physletb.2019.02.028}{\emph{Phys.\ Lett.\ B}
  {\bfseries 791} (2019) 54}
  [\href{https://arxiv.org/abs/1812.09677}{{\ttfamily 1812.09677}}].

\bibitem{Kobayashi:2018wkl}
T.~Kobayashi, Y.~Shimizu, K.~Takagi, M.~Tanimoto, T.~H. Tatsuishi and
  H.~Uchida, \emph{{Finite modular subgroups for fermion mass matrices and
  baryon/lepton number violation}},
  \href{https://doi.org/10.1016/j.physletb.2019.05.034}{\emph{Phys.\ Lett.\ B}
  {\bfseries 794} (2019) 114}
  [\href{https://arxiv.org/abs/1812.11072}{{\ttfamily 1812.11072}}].

\bibitem{Okada:2019uoy}
H.~Okada and M.~Tanimoto, \emph{{Towards unification of quark and lepton
  flavors in $A_4$ modular invariance}},
  \href{https://doi.org/10.1140/epjc/s10052-021-08845-y}{\emph{Eur. Phys. J. C}
  {\bfseries 81} (2021) 52} [\href{https://arxiv.org/abs/1905.13421}{{\ttfamily
  1905.13421}}].

\bibitem{Kobayashi:2019rzp}
T.~Kobayashi, Y.~Shimizu, K.~Takagi, M.~Tanimoto and T.~H. Tatsuishi,
  \emph{{Modular $S_3$-invariant flavor model in SU(5) grand unified theory}},
  \href{https://doi.org/10.1093/ptep/ptaa055}{\emph{PTEP} {\bfseries 2020}
  (2020) 053B05} [\href{https://arxiv.org/abs/1906.10341}{{\ttfamily
  1906.10341}}].

\bibitem{Lu:2019vgm}
J.-N. Lu, X.-G. Liu and G.-J. Ding, \emph{{Modular symmetry origin of texture
  zeros and quark lepton unification}},
  \href{https://doi.org/10.1103/PhysRevD.101.115020}{\emph{Phys. Rev. D}
  {\bfseries 101} (2020) 115020}
  [\href{https://arxiv.org/abs/1912.07573}{{\ttfamily 1912.07573}}].

\bibitem{Okada:2020rjb}
H.~Okada and M.~Tanimoto, \emph{{Quark and lepton flavors with common modulus
  $\tau$ in $A_4$ modular symmetry}},
  \href{https://arxiv.org/abs/2005.00775}{{\ttfamily 2005.00775}}.

\bibitem{Chen:2021zty}
P.~Chen, G.-J. Ding and S.~F. King, \emph{{SU(5) GUTs with A$_{4}$ modular
  symmetry}}, \href{https://doi.org/10.1007/JHEP04(2021)239}{\emph{JHEP}
  {\bfseries 04} (2021) 239}
  [\href{https://arxiv.org/abs/2101.12724}{{\ttfamily 2101.12724}}].

\bibitem{deMedeirosVarzielas:2019cyj}
I.~De~Medeiros~Varzielas, S.~F. King and Y.-L. Zhou, \emph{{Multiple modular
  symmetries as the origin of flavour}},
  \href{https://doi.org/10.1103/PhysRevD.101.055033}{\emph{Phys.\ Rev.\ D}
  {\bfseries 101} (2020) 055033}
  [\href{https://arxiv.org/abs/1906.02208}{{\ttfamily 1906.02208}}].

\bibitem{King:2019vhv}
S.~F. King and Y.-L. Zhou, \emph{{Trimaximal TM$_1$ mixing with two modular
  $S_4$ groups}},
  \href{https://doi.org/10.1103/PhysRevD.101.015001}{\emph{Phys.\ Rev.\ D}
  {\bfseries 101} (2020) 015001}
  [\href{https://arxiv.org/abs/1908.02770}{{\ttfamily 1908.02770}}].

\bibitem{Ding:2020zxw}
G.-J. Ding, F.~Feruglio and X.-G. Liu, \emph{{Automorphic Forms and Fermion
  Masses}}, \href{https://doi.org/10.1007/JHEP01(2021)037}{\emph{JHEP}
  {\bfseries 01} (2021) 037}
  [\href{https://arxiv.org/abs/2010.07952}{{\ttfamily 2010.07952}}].

\bibitem{Kobayashi:2019uyt}
T.~Kobayashi, Y.~Shimizu, K.~Takagi, M.~Tanimoto, T.~H. Tatsuishi and
  H.~Uchida, \emph{{$CP$ violation in modular invariant flavor models}},
  \href{https://doi.org/10.1103/PhysRevD.101.055046}{\emph{Phys. Rev. D}
  {\bfseries 101} (2020) 055046}
  [\href{https://arxiv.org/abs/1910.11553}{{\ttfamily 1910.11553}}].

\bibitem{Yao:2020qyy}
C.-Y. Yao, J.-N. Lu and G.-J. Ding, \emph{{Modular Invariant $A_{4}$ Models for
  Quarks and Leptons with Generalized CP Symmetry}},
  \href{https://doi.org/10.1007/JHEP05(2021)102}{\emph{JHEP} {\bfseries 05}
  (2021) 102} [\href{https://arxiv.org/abs/2012.13390}{{\ttfamily
  2012.13390}}].

\bibitem{Wang:2021mkw}
X.~Wang and S.~Zhou, \emph{{Explicit Perturbations to the Stabilizer $\tau =
  {\rm i}$ of Modular $A^\prime_5$ Symmetry and Leptonic CP Violation}},
  \href{https://doi.org/10.1007/JHEP07(2021)093}{\emph{JHEP} {\bfseries 07}
  (2021) 093} [\href{https://arxiv.org/abs/2102.04358}{{\ttfamily
  2102.04358}}].

\bibitem{Ding:2021iqp}
G.-J. Ding, F.~Feruglio and X.-G. Liu, \emph{{CP symmetry and symplectic
  modular invariance}},
  \href{https://doi.org/10.21468/SciPostPhys.10.6.133}{\emph{SciPost Phys.}
  {\bfseries 10} (2021) 133}
  [\href{https://arxiv.org/abs/2102.06716}{{\ttfamily 2102.06716}}].

\bibitem{Nilles:2020nnc}
H.~P. Nilles, S.~Ramos-S\'anchez and P.~K.~S. Vaudrevange, \emph{{Eclectic
  Flavor Groups}}, \href{https://doi.org/10.1007/JHEP02(2020)045}{\emph{JHEP}
  {\bfseries 02} (2020) 045}
  [\href{https://arxiv.org/abs/2001.01736}{{\ttfamily 2001.01736}}].

\bibitem{Kikuchi:2020nxn}
S.~Kikuchi, T.~Kobayashi, H.~Otsuka, S.~Takada and H.~Uchida, \emph{{Modular
  symmetry by orbifolding magnetized $T^2\times T^2$: realization of double
  cover of $\Gamma_N$}},
  \href{https://doi.org/10.1007/JHEP11(2020)101}{\emph{JHEP} {\bfseries 11}
  (2020) 101} [\href{https://arxiv.org/abs/2007.06188}{{\ttfamily
  2007.06188}}].

\bibitem{Liu:2019khw}
X.-G. Liu and G.-J. Ding, \emph{{Neutrino Masses and Mixing from Double
  Covering of Finite Modular Groups}},
  \href{https://doi.org/10.1007/JHEP08(2019)134}{\emph{JHEP} {\bfseries 08}
  (2019) 134} [\href{https://arxiv.org/abs/1907.01488}{{\ttfamily
  1907.01488}}].

\bibitem{Liu:2020akv}
X.-G. Liu, C.-Y. Yao and G.-J. Ding, \emph{{Modular invariant quark and lepton
  models in double covering of $S_4$ modular group}},
  \href{https://doi.org/10.1103/PhysRevD.103.056013}{\emph{Phys. Rev. D}
  {\bfseries 103} (2021) 056013}
  [\href{https://arxiv.org/abs/2006.10722}{{\ttfamily 2006.10722}}].

\bibitem{Wang:2020lxk}
X.~Wang, B.~Yu and S.~Zhou, \emph{{Double covering of the modular $A_5$ group
  and lepton flavor mixing in the minimal seesaw model}},
  \href{https://doi.org/10.1103/PhysRevD.103.076005}{\emph{Phys. Rev. D}
  {\bfseries 103} (2021) 076005}
  [\href{https://arxiv.org/abs/2010.10159}{{\ttfamily 2010.10159}}].

\bibitem{Yao:2020zml}
C.-Y. Yao, X.-G. Liu and G.-J. Ding, \emph{{Fermion masses and mixing from the
  double cover and metaplectic cover of the $A_5$ modular group}},
  \href{https://doi.org/10.1103/PhysRevD.103.095013}{\emph{Phys. Rev. D}
  {\bfseries 103} (2021) 095013}
  [\href{https://arxiv.org/abs/2011.03501}{{\ttfamily 2011.03501}}].

\bibitem{Liu:2021gwa}
X.-G. Liu and G.-J. Ding, \emph{{Modular flavor symmetry and vector-valued
  modular forms}},  \href{https://arxiv.org/abs/2112.14761}{{\ttfamily
  2112.14761}}.

\bibitem{Kobayashi:2018rad}
T.~Kobayashi, S.~Nagamoto, S.~Takada, S.~Tamba and T.~H. Tatsuishi,
  \emph{{Modular symmetry and non-Abelian discrete flavor symmetries in string
  compactification}},
  \href{https://doi.org/10.1103/PhysRevD.97.116002}{\emph{Phys.\ Rev.\ D}
  {\bfseries 97} (2018) 116002}
  [\href{https://arxiv.org/abs/1804.06644}{{\ttfamily 1804.06644}}].

\bibitem{Kobayashi:2020hoc}
T.~Kobayashi and H.~Otsuka, \emph{{Classification of discrete modular
  symmetries in Type IIB flux vacua}},
  \href{https://doi.org/10.1103/PhysRevD.101.106017}{\emph{Phys. Rev. D}
  {\bfseries 101} (2020) 106017}
  [\href{https://arxiv.org/abs/2001.07972}{{\ttfamily 2001.07972}}].

\bibitem{Abe:2020vmv}
H.~Abe, T.~Kobayashi, S.~Uemura and J.~Yamamoto, \emph{{Loop Fayet-Iliopoulos
  terms in $T^2/Z_2$ models: Instability and moduli stabilization}},
  \href{https://doi.org/10.1103/PhysRevD.102.045005}{\emph{Phys. Rev. D}
  {\bfseries 102} (2020) 045005}
  [\href{https://arxiv.org/abs/2003.03512}{{\ttfamily 2003.03512}}].

\bibitem{Ohki:2020bpo}
H.~Ohki, S.~Uemura and R.~Watanabe, \emph{{Modular flavor symmetry on a
  magnetized torus}},
  \href{https://doi.org/10.1103/PhysRevD.102.085008}{\emph{Phys. Rev. D}
  {\bfseries 102} (2020) 085008}
  [\href{https://arxiv.org/abs/2003.04174}{{\ttfamily 2003.04174}}].

\bibitem{Nilles:2020kgo}
H.~P. Nilles, S.~Ramos-Sanchez and P.~K.~S. Vaudrevange, \emph{{Lessons from
  eclectic flavor symmetries}},
  \href{https://doi.org/10.1016/j.nuclphysb.2020.115098}{\emph{Nucl. Phys. B}
  {\bfseries 957} (2020) 115098}
  [\href{https://arxiv.org/abs/2004.05200}{{\ttfamily 2004.05200}}].

\bibitem{Nilles:2020tdp}
H.~P. Nilles, S.~Ramos\textendash{}S\'anchez and P.~K.~S. Vaudrevange,
  \emph{{Eclectic flavor scheme from ten-dimensional string theory
  \textendash{} I. Basic results}},
  \href{https://doi.org/10.1016/j.physletb.2020.135615}{\emph{Phys. Lett. B}
  {\bfseries 808} (2020) 135615}
  [\href{https://arxiv.org/abs/2006.03059}{{\ttfamily 2006.03059}}].

\bibitem{Ishiguro:2020nuf}
K.~Ishiguro, T.~Kobayashi and H.~Otsuka, \emph{{Spontaneous CP violation and
  symplectic modular symmetry in Calabi-Yau compactifications}},
  \href{https://doi.org/10.1016/j.nuclphysb.2021.115598}{\emph{Nucl. Phys. B}
  {\bfseries 973} (2021) 115598}
  [\href{https://arxiv.org/abs/2010.10782}{{\ttfamily 2010.10782}}].

\bibitem{Ishiguro:2020tmo}
K.~Ishiguro, T.~Kobayashi and H.~Otsuka, \emph{{Landscape of Modular Symmetric
  Flavor Models}}, \href{https://doi.org/10.1007/JHEP03(2021)161}{\emph{JHEP}
  {\bfseries 03} (2021) 161}
  [\href{https://arxiv.org/abs/2011.09154}{{\ttfamily 2011.09154}}].

\bibitem{Baur:2020yjl}
A.~Baur, M.~Kade, H.~P. Nilles, S.~Ramos-Sanchez and P.~K.~S. Vaudrevange,
  \emph{{Siegel modular flavor group and CP from string theory}},
  \href{https://doi.org/10.1016/j.physletb.2021.136176}{\emph{Phys. Lett. B}
  {\bfseries 816} (2021) 136176}
  [\href{https://arxiv.org/abs/2012.09586}{{\ttfamily 2012.09586}}].

\bibitem{Almumin:2021fbk}
Y.~Almumin, M.-C. Chen, V.~Knapp-P\'erez, S.~Ramos-S\'anchez, M.~Ratz and
  S.~Shukla, \emph{{Metaplectic Flavor Symmetries from Magnetized Tori}},
  \href{https://doi.org/10.1007/JHEP05(2021)078}{\emph{JHEP} {\bfseries 05}
  (2021) 078} [\href{https://arxiv.org/abs/2102.11286}{{\ttfamily
  2102.11286}}].

\bibitem{Baur:2021bly}
A.~Baur, H.~P. Nilles, S.~Ramos-Sanchez, A.~Trautner and P.~K.~S. Vaudrevange,
  \emph{{Top-Down Anatomy of Flavor Symmetry Breakdown}},
  \href{https://arxiv.org/abs/2112.06940}{{\ttfamily 2112.06940}}.

\bibitem{Okada:2020ukr}
H.~Okada and M.~Tanimoto, \emph{{Modular invariant flavor model of $A_4$ and
  hierarchical structures at nearby fixed points}},
  \href{https://doi.org/10.1103/PhysRevD.103.015005}{\emph{Phys. Rev. D}
  {\bfseries 103} (2021) 015005}
  [\href{https://arxiv.org/abs/2009.14242}{{\ttfamily 2009.14242}}].

\bibitem{Bailin:1997fh}
D.~Bailin, G.~V. Kraniotis and A.~Love, \emph{{CP violation by soft
  supersymmetry breaking terms in orbifold compactifications}},
  \href{https://doi.org/10.1016/S0370-2693(97)01179-9}{\emph{Phys. Lett. B}
  {\bfseries 414} (1997) 269}
  [\href{https://arxiv.org/abs/hep-th/9705244}{{\ttfamily hep-th/9705244}}].

\bibitem{Ferrara:1989bc}
S.~Ferrara, D.~Lust, A.~D. Shapere and S.~Theisen, \emph{{Modular Invariance in
  Supersymmetric Field Theories}},
  \href{https://doi.org/10.1016/0370-2693(89)90583-2}{\emph{Phys. Lett. B}
  {\bfseries 225} (1989) 363}.

\bibitem{Ferrara:1989qb}
S.~Ferrara, D.~Lust and S.~Theisen, \emph{{Target Space Modular Invariance and
  Low-Energy Couplings in Orbifold Compactifications}},
  \href{https://doi.org/10.1016/0370-2693(89)90631-X}{\emph{Phys. Lett.}
  {\bfseries B233} (1989) 147}.

\bibitem{Chen:2019ewa}
M.-C. Chen, S.~Ramos-Sánchez and M.~Ratz, \emph{{A note on the predictions of
  models with modular flavor symmetries}},
  \href{https://doi.org/10.1016/j.physletb.2019.135153}{\emph{Phys. Lett.}
  {\bfseries B801} (2020) 135153}
  [\href{https://arxiv.org/abs/1909.06910}{{\ttfamily 1909.06910}}].

\bibitem{Chen:2021prl}
M.-C. Chen, V.~Knapp-Perez, M.~Ramos-Hamud, S.~Ramos-Sanchez, M.~Ratz and
  S.~Shukla, \emph{{Quasi-Eclectic Modular Flavor Symmetries}},
  \href{https://arxiv.org/abs/2108.02240}{{\ttfamily 2108.02240}}.

\bibitem{Gonzalo:2018guu}
E.~Gonzalo, L.~E. Ib\'a\~nez and A.~M. Uranga, \emph{{Modular symmetries and
  the swampland conjectures}},
  \href{https://doi.org/10.1007/JHEP05(2019)105}{\emph{JHEP} {\bfseries 05}
  (2019) 105} [\href{https://arxiv.org/abs/1812.06520}{{\ttfamily
  1812.06520}}].

\bibitem{Cvetic:1991qm}
M.~Cvetic, A.~Font, L.~E. Ibanez, D.~Lust and F.~Quevedo, \emph{{Target space
  duality, supersymmetry breaking and the stability of classical string
  vacua}}, \href{https://doi.org/10.1016/0550-3213(91)90622-5}{\emph{Nucl.
  Phys. B} {\bfseries 361} (1991) 194}.

\bibitem{Ibanez:2012zz}
L.~E. Ibanez and A.~M. Uranga, \emph{{String theory and particle physics: An
  introduction to string phenomenology}}. Cambridge University Press, 2, 2012.

\bibitem{Feruglio:2021dte}
F.~Feruglio, V.~Gherardi, A.~Romanino and A.~Titov, \emph{{Modular invariant
  dynamics and fermion mass hierarchies around $\tau = i$}},
  \href{https://doi.org/10.1007/JHEP05(2021)242}{\emph{JHEP} {\bfseries 05}
  (2021) 242} [\href{https://arxiv.org/abs/2101.08718}{{\ttfamily
  2101.08718}}].

\bibitem{Schoeneberg:1974md}
B.~Schoeneberg, \emph{Elliptic Modular Functions: An Introduction}, Grundlehren
  der mathematischen Wissenschaften. Springer, Berlin, Heidelberg, 1974.

\bibitem{Apostol:1990mf}
T.~M. Apostol, \emph{Modular Functions and Dirichlet Series in Number Theory},
  Graduate Texts in Mathematics. Springer, New York, NY, 1990.

\end{thebibliography}\endgroup

\end{document}